\documentclass [12pt,a4wide]{article}
\usepackage{times}
\usepackage{amsmath,amsthm,amsfonts,amscd,eucal,latexsym,amssymb,mathrsfs,cancel}
\usepackage{epsfig}  
\usepackage{simplewick,bm}
\usepackage{hyperref}
\usepackage{hyperref}
\usepackage{tikz}
\input xy
\xyoption{all}
\usepackage{xcolor}
 
\def\beq{\begin{eqnarray}}
\def\eeq{\end{eqnarray}}

\def\ka{\kappa}

\def\sv{v_*}
\def\sr{r_*}
\def\stheta{\vartheta_*}
\def\sphi{\varphi_*}

\newcommand{\nn}{\boldsymbol{\nu}}

\newcommand{\mpof}{{\mbox{\small M}}}

\newtheorem{theorem}{Theorem}[section]

\newtheorem{proposition}{Proposition}[section]
\newtheorem{lemma}{Lemma}[section]
\theoremstyle{definition}

\begin{document}

%
%
 
\par 
\bigskip 
\LARGE 
\noindent 
{\bf Temperature and entropy-area relation of quantum matter near spherically symmetric outer trapping horizons} 
\bigskip 
\par 
\rm 
\normalsize 
 

\large
\noindent 
{\bf Fiona Kurpicz$^{1,a}$}, {\bf Nicola Pinamonti$^{2,3,b}$}, {\bf Rainer Verch$^{1,c}$} \\
\par
\small

\noindent$^1$ Institute for Theoretical Physics, University of Leipzig, D-04009 Leipzig, Germany.\smallskip

\noindent$^2$ Dipartimento di Matematica, Universit\`a di Genova - Via Dodecaneso, 35, I-16146 Genova, Italy. \smallskip

\noindent$^3$ Istituto Nazionale di Fisica Nucleare - Sezione di Genova, Via Dodecaneso, 33 I-16146 Genova, Italy. \smallskip
\smallskip

\noindent E-mail: 
$^a$kurpicz@itp.uni-leipzig.de , 
$^b$pinamont@dima.unige.it,
$^c$Rainer.Verch@uni-leipzig.de\\ 

\normalsize

\par 
 
\rm\normalsize 

 
 
\par

\rm\normalsize

\bigskip

\noindent 
\small 
{\bf Abstract} 
\\[6pt]
We consider spherically symmetric spacetimes with an outer trapping horizon. Such spacetimes are generalizations of spherically symmetric black hole spacetimes
where the central mass can vary with time, like in black hole collapse or black hole evaporation. While these spacetimes possess in general no timelike 
Killing vector field, they admit a Kodama vector field which in some ways provides a replacement. The Kodama vector field allows the definition of a surface gravity 
of the outer trapping horizon. Spherically symmetric spacelike cross-sections of the outer trapping horizon define in- and outgoing lightlike congruences.
We investigate a scaling limit of Hadamard 2-point functions of a quantum field on the spacetime onto the ingoing lightlike congruence. The scaling limit 
2-point function has a universal form and a thermal spectrum with respect to the time-parameter of the Kodama flow, where the inverse temperature 
$\beta = 2\pi/\kappa$ is related to the surface gravity $\kappa$  of the horizon cross-section in the same way as in the Hawking effect for an asymptotically 
static black hole. Similarly, the tunneling probability that can be obtained in the scaling limit between in- and outgoing Fourier modes with respect to 
the time-parameter of the Kodama flow shows a thermal distribution with the same inverse temperature, determined by the surface gravity. This can be seen as a
local counterpart of the Hawking effect for a dynamical horizon in the scaling limit. Moreover, the scaling limit 2-point function allows it to define 
a scaling-limit-theory, a quantum field theory on the ingoing lightlike congruence emanating from a horizon cross-section. The scaling limit 2-point function
as well as the 2-point functions of coherent states of the scaling-limit-theory are correlation-free with respect to separation along the horizon-cross section,
therefore, their relative entropies behave proportional to the cross-sectional area. We thus obtain a proportionality of the relative entropy of coherent states of the 
scaling-limit-theory and the area of the horizon cross-section with respect to which the scaling limit is defined. Thereby, we establish a local counterpart, and 
microscopic interpretation in
the setting of quantum field theory on curved spacetimes, of the dynamical laws of outer trapping horizons, derived by Hayward and others in generalizing the 
laws of black hole dynamics originally shown for stationary black holes by Bardeen, Carter and Hawking.

%
\normalsize

\vskip .3cm


\section{Introduction}

The famous four laws of black hole mechanics and their analogy with the laws of thermodynamics have been derived and developed in
\cite{BardeenCarterHawking} assuming stationarity.
The temperature thereby assigned to a black hole is related to the horizon's surface gravity and can physically be interpreted in terms of 
Hawking radiation \cite{Hawking} in the framework of quantum field theory in curved spacetime, see also \cite{FredenhagenHaag,KayWald,Sewell,Wald1975}. Similarly, the area of the black hole horizon 
surface is analogous to an entropy. Discussions of black hole entropy and its physical nature have been given in a variety of contexts (\cite{Bekenstein,WaldTDBH,FabbriNavarro,SoloLRR,Perez,MannBH} and literature 
cited therein is just a small sample of references on the topic) but it has been difficult 
to find a simple, direct counterpart of the entropy-area relation for black holes in the setting of quantum field theory in curved spacetime (see however \cite{HollIshi}, and further discussion below). 

Although Hawking radiation is derived neglecting backreaction, assuming that the spacetime geometry is stationary
(or asymptotically stationary), 
the emission rate of Hawking radiation is usually associated to the rate of black hole mass loss due to evaporation, see e.g.\ 
\cite{Hawking,Candelas,FabbriNavarro}.
However, black hole evaporation is a dynamical process and should be described locally. 
A local theory for the geometry of non-stationary black holes using concepts of dynamical horizons
and trapped horizons has been developed, see e.g.\ \cite{Hayward, Ashtekar}.
In particular, in \cite{Hayward} it is shown that the first law holds as an energy balance along the trapped horizon.  
In contrast to the (asymptotically) stationary case, Hawking radiation and a relation between temperature
and local geometrical quantities of dynamical or trapped horizons has so far not been derived for quantum fields
in the background of {\it non-stationary (or dynamical)}
black holes.
\par
An essentially local derivation of the Hawking effect has been proposed by Parikh and Wilczeck \cite{PW}.
In that approach, an estimate is given for the tunneling probability 
of quantum particles across the horizon, showing that this probability has a thermal distribution.
This idea has been generalized to the case of dynamical black holes in \cite{Zerbini1,Zerbini2,Zerbini3},
hence furnishing a connection between the surface gravity and a thermal distribution of the tunneling probability.
These considerations didn't use quantum field theoretical methods as in the original derivation of the Hawking effect
but relied on single-particle quantum mechanics in a WKB-type approximation. In order to overcome the limitations of such a quantum 
mechanical treatment, it has been shown in \cite{MP} that for a scalar quantum field on a stationary black hole 
spacetime --- more generally,
any spacetime with a bifurcate Killing horizon --- a thermal distribution in the tunneling probability is obtained 
in a certain scaling limit located on the horizon whenever the quantum field is in a Hadamard state. The associated temperature is 
the Hawking temperature and is independent of the chosen Hadamard state. A similar result can also be obtained in the case of 
self-interacting fields, see \cite{Collini}.
For some related results, focusing on the thermal nature of field theories restricted on null surfaces (horizons) and thus not focussing on the local 
aspect related to tunneling processes, see  \cite{Sewell, KayWald, GLRV, FredenhagenHaag, HaagNarnhoferStein, SummersVerch}.
\par
In this paper we aim at generalizing the result of \cite{MP} to
the case of spherically symmetric, dynamical black holes.
For generic spherically symmetric black holes there is  no Killing vector field which generates an horizon.
 Nevertheless, there are generalizations available that serve a similar purpose in the context of black hole thermodynamics.
In particular, we shall use the concept of {\it outer trapping horizons} \cite{Ashtekar} and the {\it Kodama vector field} \cite{Kodama}. 
 A spherically symmetric (non-stationary)  black hole spacetime is a warped product of a 2-dimensional Lorentzian space 
and a 2-dimensional Euclidean sphere. The future-directed light rays in the two-dimensional Lorentzian space determine, at each spacetime point,
two geodesic congruences of null type;
one is called {\it outgoing} and the other {\it ingoing}. The {\it outer trapping horizon} $\mathcal{H}$ is the 3-dimensional 
hypersurface which divides the {\it inside} region where the expansion parameters $\theta_\pm$ 
of the ingoing$(-)$ and outgoing$(+)$ null geodesic congruences are both negative from 
the {\it outside} region where $\theta_+>0$ and $\theta_-<0$. The outside region usually reaches out to spatial infinity.
If the expansion parameter of the null geodesic congruence is positive (negative),
the area of a congruence-orthogonal spatial sphere grows (decreases) towards the future along the congruence.
Hence, in the region where both $\theta_\pm$ are negative all light rays tend to fall into the black hole while in the
region where $\theta_+>0$ the outgoing lightrays tend to reach points which are far away 
(measured by the radius of the orthogonal spatial sphere)
from the center of the black hole.
Thus, an outer trapping horizon $\mathcal{H}$ is the surface from which nothing can escape instantaneously. It is
worth noting that in a dynamical spherically symmetric spacetime, $\mathcal{H}$ need not be lightlike but
can have timelike or spacelike parts. 
\par
In \cite{Kodama}, Kodama has shown that in the
case of spherically symmetric spacetimes,
it is possible to find a vector field\footnote{We shall mostly employ the abstract index notation for vector and tensor fields
as in \cite{Wald} } 
$K^a$ which can be used
as a replacement of the timelike Killing vector field of a stationary black hole (the full definition will be given
in Sec. \ref{se:outer-trapping-horizon}). 
This {\it Kodama vector field} $K^a$
is a conserved current, and also $W^{ab}K_b$ is a conserved current whenever $W^{ab}$ is a symmetric tensor field that 
is invariant under the spherical symmetries of the spacetime. 
Furthermore, $K^a$ is timelike outside of, spacelike inside of, and lightlike on an outer trapping horizon $\mathcal{H}$, respectively.
On  $\mathcal{H}$, one has 
\begin{equation}\label{eq:Intro-kappa}
\frac{1}{2}K^a (\nabla_a K_b-\nabla_b K_a) = \kappa K_b  
\end{equation}
where the function $\kappa$ is the {\it surface gravity} along $\mathcal{H}$.
\par
Outer trapping horizons and the conservation of currents generated by the Kodama vector field have been used by 
Hayward \cite{Hayward} to derive a first thermodynamical law for dynamical black holes.  
In particular, it holds that 
\[
\mathcal{M}' = \frac{\kappa }{8\pi} \mathcal{A}' + w \mathcal{V}'
\]
with the derivative $f' =  z^a \nabla_a f$ where $z^a$ is any (nowhere vanishing) vector 
field having zero angular components tangent to the outer trapping horizon.
Furthermore, $\mathcal{M}$ is the Hawking mass of the black hole, $\mathcal{A}=4\pi r^2$ is the area of the 
surface, $\mathcal{V} = \frac{4}{3}\pi r^3$ is the surface-enclosed volume, and
$\kappa$ is the surface gravity associated to the Kodama vector field.
The term $w = - G_{UV}g^{UV}$ is related to the trace of the Einstein tensor 
taken with respect to the lightlike coordinates of the horizon, symbolized by indices $U$ and $V$;
see Section \ref{se:geometry} for full details.
 As usual, $\mathcal{M}$  is interpreted as the black hole's internal energy, see \eqref{eq:Hawking-mass} below, and
 $w\mathcal{V}'$ is the work done on the system. Interpreting $\kappa/(2\pi)$  as a temperature, 
 $\mathcal{A}'/4$ represents the variation of entropy. 
\par
 We will consider a (for simplicity,
 scalar) quantum field $\phi(x)$ propagating on a spherically symmetric spacetime with an outer trapping horizon and
 a Kodama vector field. Here, we follow common practice to write symbolically $\phi(x)$ where $x$ is a spacetime point
as if $\phi(x)$ was an operator-valued function, while actually it is an operator-valued distribution. We will take 
due care of this circumstance whenever required in the main body of the text. To further simplify matters, we assume that $\phi(x)$ is a quantized Klein-Gordon field fulfilling the field equation 
$(\nabla^a\nabla_a - \mpof(x))\phi(x) = 0$  where $\nabla$ is the covariant derivative
of the spacetime metric $g_{ab}$ and $\mpof$ is a smooth, real-valued function on spacetime. 
(This assumption could, in fact, be generalized.)
\par
States (and in particular, quasifree states) of the quantized Klein-Gordon field on curved spacetimes
admitting a physical interpretation consistent with the principles that 
apply for quantum field theory on Minkowski spacetime are {\it Hadamard states}.
 These states are 
defined as having a 2-point function of {\it Hadamard form}, meaning that 
\begin{equation}\label{eq:Intro-HadForm}
w^{(2)}(x_1,x_1) =   
\frac{1}{8\pi^2}\frac{{\Delta}^{1/2}(x_1,x_2)}{\sigma_\epsilon(x_1,x_2)} + W_\epsilon(x_1,x_2),
\end{equation}
where ${\Delta}$ is the {\it van Vleck\,-\,Morette determinant} of the 
spacetime metric and 
${\sigma}(x_1,x_2)$ is its Synge function, i.e.\ the squared geodesic distance divided by 2. 
Both quantities are determined by the spacetime metric;
the subscript $\epsilon$ denotes a regularisation that is used to properly define the quantity on the 
right hand side as a distribution (after integration with test functions) in the limit $\epsilon \to 0$
(see \cite{KayWald}  and Sec.\ \ref{se:QF} for further details).
Similarly, in the limit $\epsilon \to 0$, $W_\epsilon(x_1,x_2)$ 
is a distribution which diverges at most logarithmically in ${\sigma}$ for ${\sigma}\to 0$
and contains the state-dependence as a smooth contribution.
For a discussion as to why Hadamard states are of particular significance, see
e.g.\ \cite{FewsterVerch,KhavkineMoretti,Wald2} and references cited there.
\par
\begin{figure}
    \centering
    \begin{tikzpicture}[scale=2.75, every node/.style={transform shape}]
        \draw [black] (-0.9,-0.9) -- (0.9,0.9);                                           
        \draw [black] (0.8,0.87) -- (0.9,0.9) -- (0.87,0.8);                              
        \node [black, below right, scale=0.36] at (0.9,0.9) {$V$};                         
        \node [black,  scale=0.36] at (0.6,0.4) {$\mathcal{C}_*$};               
        \draw [black] (0.9,-0.9) -- (-0.9,0.9);                                          
        \draw [black] (-0.8,0.87) -- (-0.9,0.9) -- (-0.87,0.8);                          
        \node [black, below left, scale=0.36] at (-0.9,0.9) {$U$};                        
        \node [black, scale=0.36] at (-0.6,0.4) {$\mathcal{T}_*$};            
        \draw [thick, domain=-0.6:0.6, samples=100] plot ({\x*cos(35) + sin(35)*(2*\x+0.05*sin(25*\x*\x r)+(\x-0.6)*(\x-0.4)*(\x-0.2)+exp(-\x)-1+0.6*0.4*0.2+0.6*\x*\x)},{ -\x*sin(35) + cos(35)*(2*\x+0.05*sin(25*\x*\x r)+(\x-0.6)*(\x-0.4)*(\x-0.2)+exp(-\x)-1+0.6*0.4*0.2+0.6*\x*\x)});            
        \node [black, scale=0.36] at (0.7,0.15) {$\mathcal{H}$};                         
        \draw [fill] (0,0) circle [radius=0.03];                                        
        \node [right, scale=0.36] at (-0.13,-0.15) {$S_*$};                                   
        \begin{scope}[shift={(-0.45,-0.9)}]                                             
        \draw [black, fill=black] (0,0) -- (0.1, 0.1) -- (-0.1, 0.1) -- (0,0);          
        \draw [black]             (0,0) -- (0.1,-0.1) -- (-0.1,-0.1) -- (0,0);          
        \end{scope}
        \draw [opacity=0] (-1.1,-1.1) -- (-1.1,1.1) -- (1.1,1.1) -- (1.1,-1.1) -- cycle;
    \end{tikzpicture}
    \caption{The picture represents the $U,V$ plane in adapted coordinates $(U,V,\vartheta,\varphi)$. The thick line corresponds 
    to the outer trapping horizon $\mathcal{H}$, $S_*\subset \mathcal{H}$ is a sphere which is used to identify the null congruence $\mathcal{C}_*$ 
    towards which we compute the scaling limit of the quantum states. The scaling limit state is then restricted onto the null congruence $\mathcal{T}_*$.}
    \label{fig1}
\end{figure}
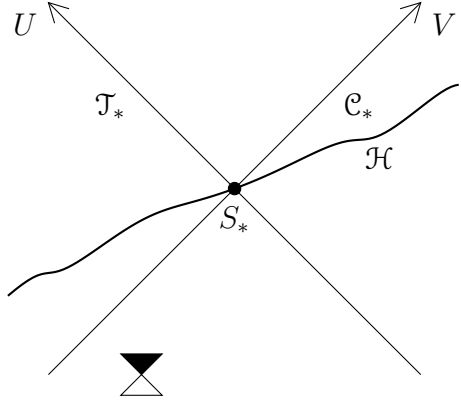

We will show that close to an outer trapping horizon of a spherically symmetric spacetime, the universal leading
short-distance singularity behaviour of any Hadamard state results, in a scaling limit, in an interpretation
of the surface gravity $\kappa$ as a temperature parameter, in close analogy to previous considerations 
for the case of quantum fields on stationary black holes \cite{KayWald,Hollands,MP,SummersVerch,GLRV}. 
Our approach follows the spirit of \cite{MP} very closely and thus makes contact with the tunneling 
interpretation of Hawking radiation. For a spherically symmetric spacetime with outer trapping horizon 
$\mathcal{H}$, one introduces Eddington-Finkelstein coordinates $v,r,\vartheta,\varphi$. Some point on $\mathcal{H}$
will be determined by certain coordinate values $(\sv,\sr,\stheta,\sphi)$ and, by spherical symmetry, it determines
the associated spatial spherical cross-section $S_* = S(\sv,\sr)$ of $\mathcal{H}$. The outgoing null geodesic congruence 
emanating from $S_*$ defines a null hypersurface denoted by $\mathcal{C}_*$, and similarly the ingoing null
geodesic congruence emanating from $S_*$ defines a null hypersurface denoted by $\mathcal{T}_*$.
As will be discussed in the main body of this paper, given $S_*$, there exists 
a natural choice of an affine parameter $V$ along the geodesic generators of $\mathcal{C}_*$ and of an affine parameter
$U$ along the geodesic generators of $\mathcal{T}_*$ so that local coordinates $(U,V,\vartheta,\varphi)$ near
$S_*$ can be introduced, with the following properties:
\\[2pt]
${}$ \quad 
(1) $U = 0$ and $V = 0$ exactly for the points on $S_*$, 
\\[2pt]
${}$ \quad 
(2) 
$U = 0$ exactly for the points on $\mathcal{C}_*$, 
\\[2pt]
${}$ \quad 
(3) $V = 0$ exactly for the points on $\mathcal{T}_*$, 
\\[2pt]
${}$ \quad 
(4)
$ds^2 = -2A(U,V)dU\,dV + r^2(U,V) d\Omega^2$ is the metric line element where $d\Omega^2$ denotes\\
${}$ \quad \ \
the line element of
the two-dimensional Euclidean sphere, and $A = 1$ on $\mathcal{C}_* \cup \mathcal{T}_*$, 
\\[2pt]
${}$ \quad 
(5) $dU_a K^a (U,V = 0,\vartheta,\varphi) = -\kappa_* U + O(U^2)$
on $\mathcal{T}_*$ near $U = 0$, with $\kappa_* = \left. \kappa\right|_{S_*}$.
\\[2pt]
We call $(U,V,\vartheta,\varphi)$ with the properties stated above \emph{adapted coordinates} with respect to $S_*$ (See Fig.\ \ref{fig1} for an illustration.)
\\[2pt] ${}$
 To analyze the short distance behavior of the 2-point function of Hadamard states when both $x_1$ and $x_2$ are very close to $\mathcal{H}$ we proceed as follows. 
 Once a sphere $S_*$ (having radius $r_*$) of the outer trapping horizon is chosen and the null surface $\mathcal{C}_*$ of outgoing null geodesics is determined, 
 we take a suitable scaling limit of the 2-point function towards $\mathcal{C}_*$. As we shall prove in Theorem \ref{Thm:SL}, the 2-point function (distribution) $\Lambda$ 
 thus obtained
 is universal, and 
it can be tested with compactly supported smooth functions on $\mathcal{T}_*$. Using adapted coordinates, its 
regularized integral kernel has the form
\begin{align}
\Lambda_\varepsilon(U,\nn; U',\nn') =  -\frac{1}{{\pi}}\frac{r_*^2}{(U-U'+i\epsilon)^2}\delta(\nn,\nn'),
\end{align} 
where $(U,\nn)$ denotes a point on $\mathcal{T}_*$, $U$ is the null coordinate and $\nn =(\vartheta, \varphi)$ denotes standard angular coordinates on the sphere $S_*$.
 Furthermore $\varepsilon > 0$ is a regulator (to be taken to 0 after 
integrating against test functions) and $\delta(\nn,\nn')$ is the Dirac 
delta function supported on coinciding angles.\footnote{Formally, this means that 
$f(\nn) = \int_{S^2} \delta(\nn,\nn') f(\nn') \,d\Omega^2(\nn')$ for any continuous function $f$ on the unit sphere.}
As we shall see in Sec.\ \ref{se:thermality}, the thermal properties are manifest   
when $\Lambda$ is tested with respect to the flow $\Phi_\tau$ ($\tau \in \mathbb{R})$ generated by $K^a$. 
Applied to $\Lambda$ the flow acts as  $\Phi_\tau(U) = e^{\ka_* \tau} U$ where $\ka_*$ is the surface gravity on $S_*$, and 
the Fourier frequencies, or energies, with respect to the flow-parameter $\tau$ in the spectrum of $\Lambda$
are distributed according to the spectral density
\begin{align}
\rho(E) = \frac{E}{1-e^{-\frac{2\pi}{\ka_*} E}}\,.
\end{align}
The presence of a Bose factor with inverse temperature $2\pi/\ka_*$ in the spectral density distribution makes the thermal interpretation manifest, analogously as in \cite{MP}.
Making use of this fact, in following \cite{MP} we  show that the tunneling probability, or transition probability, between a one-particle state inside the outer trapping horizon $\mathcal{H}$, i.e.\ for 
$U > 0$ and another one particle state outside of $\mathcal{H}$, i.e.\ for $U<0$, takes in the scaling limit the high-energy asymptotic form ${\rm e}^{-\beta E}$, when the 
the one-particle states have a Fourier distribution peaked at $E$. This is the form of a transition probability for a thermal energy level occupation at inverse temperature $\beta = 2 \pi/\kappa_*$.

The 2-point function $\Lambda$ obtained in our scaling limit is very similar to the restriction  of 2-point functions to Killing horizons considered in  \cite{Sewell,HaagNarnhoferStein,KayWald,SummersVerch,GLRV}.
In these articles, the restrictions or scaling limits of 2-point functions to the analogues of $\mathcal{C_*}$ exhibit a thermal spectrum with respect to the Killing flow. In contrast, in the case of  dynamical black holes the relevant 
part of the state is the transversal component of the 2-point function (the component supported on $\mathcal{T}_*$), showing thermal properties with respect to the Kodama flow in the scaling limit. 
The $\mathcal{C}_*$-part of the 2-point function depends on the details of the quantum matter entering the horizon, blurring an exact thermal spectrum. 
On the other hand, at least in the case of static black holes, the $\mathcal{T}_*$-part is related to the radiation emitted by the black hole and is the source of
Hawking radiation, see e.g.\ \cite{FredenhagenHaag}. 

The 2-point function $\Lambda$ can be used to define a quantum field theory -- the ``scaling-limit-theory'' -- on the lightlight hypersurface $\mathcal{T_*}$; $\Lambda$ also induces a quasifree state $\omega_\Lambda$ 
on the algebra of observables $\mathcal{W}$ of the scaling-limit-theory, where $\mathcal{W}$ is a CCR-Weyl algebra. 
This state turns out
to be a KMS-state \cite{KMS} at inverse temperature $\beta = 2\pi/\kappa_*$ with respect to the Kodama flow.
It is then possible to define and calculate
the relative entropy $S(\omega_\Lambda|\omega_\varphi)$ in the sense of Araki \cite{Araki1} 
between $\omega_\Lambda$ and coherent states $\omega_\varphi$ on $\mathcal{W}$ analogously as in \cite{HollandsRECS,LongoECE}
where the fucntion $\varphi$ describes a coherent excitation of the scalar field over the state $\omega_\Lambda$. 
We find that $S(\omega_\Lambda|\omega_\varphi)$ coincides with the classical energy of the coherent excitation, measured by an observer moving along the Kodama flow, multiplied with
the inverse temperature $\beta = 2\pi/\kappa_*$ (see equation \eqref{eq:entropy-formula} below). Furthermore, we observe that 
$S(\omega_\Lambda|\omega_\varphi)$ is proportional to 
$r_*^2$, the geometrical area of the outer trapping horizon's cross section $S_*$ with respect to which the scaling-limit-theory is constructed. We argue that this is not accidental but a consequence of the 
fact that $\omega_\Lambda$ and $\omega_\varphi$ are correlation-free product states with respect to separation in the angular coordinate $\nn$ of $S_*$, together with the additivity of the relative 
entropy for correlation-free product states. Thus we arrive at $S(\omega_\Lambda | \omega_\varphi) \sim r_*^2$, analogous to the entropy-area relation suggested in the classic article of Bardeen, Carter and Hawking
\cite{BardeenCarterHawking}. 

The idea to relate the relative entropy of quantum field states on a spacetime containing a horizon to a form of black hole entropy goes back to Longo \cite{LongoBHCE} (in the setting of 
quantum field theory on Minkowski spacetime, where the lightlike boundaries of a wedge-region play the role of a horizon). These ideas have been extended in 
\cite{kl} to a relation between relative entropy of quantum field states and 
non-commutative geometrical quantities for area. In a series of articles by Schroer \cite{SchroerADOE,SchroerARLOC} it was mentioned that -- in the setting of quantum field theory in Minkowski spacetime --
quantum field theories restricted to lightlike hyperplanes typically show no correlations in the transversal spacelike directions of the hyperplane, which would result in an additive behaviour of 
entropy quantities for the restricted quantum fields and an area proportionality. In our present article, we see that the earlier ideas of Longo and of Schroer can indeed be combined to result, in our scaling limit, in a 
proportionality between the relative entropy of quantum field states and the horizon area of a black hole spacetime, and more generally, the cross-section area of an outer trapping horizon. 
In a recent work, Hollands and Ishibashi \cite{HollIshi} consider linearized perturbations of the spacetime metric around Schwarzschild spacetime which are quantized similarly like a linear scalar field.
Using preferred states for the characteristic data of the perturbations (which on the black hole horizon take a form as our scaling limit state) they  
define relative entropies in a similar way, and taking into account the backreaction of the coherent excitation on the background geometry, 
they show that the combined variation of the relative entropy and a cross-sectional area of the black-hole horizon along Schwarzschild time equals the future out- and ingoing flux of 
radiation. Related content, in the context of spherically symmetric dynamical black holes, appears also in \cite{dangelo}. 
\par
This article is organized as follows.
In Section 2 we discuss the geometric setup of spherically symmetric spacetimes in which a Kodama vector field and outer trapping horizons can be defined. 
Section 3 contains the specification of the quantum field theory on the spacetimes we consider, together with a discussion about Hadamard 2-point function.
Section 4 begins by introducting a conformal transformation of the spacetimes considered which is useful for deriving the scaling limit of Hadamard 2-point functions,
presented thereafter in Theorem 4.1. The behaviour of the scaling limit 2-point function $\Lambda$ under the Kodama flow is also discussed. 
In Section 5 we show how the scaling-limit-theory is constructed from the scaling limit 2-point function $\Lambda$, and derive and discuss several of its properties,
like the thermal spectrum and thermal tunneling probability with respect to the Kodama time. We also consider the coherent states in the scaling-limit-theory and 
their relative entropy which we find to be proportional to the area of the horizon cross-section $S_*$ with respect to which the scaling-limit-theory is defined. 
A conclusion is given in Section 6. Section 7 is a technical appendix containing the proof of Theorem 4.1.

\section{Geometric setup}\label{se:geometry}

\subsection{Spherically symmetric spacetime, Eddington-Finkelstein coordinates}

We consider a spacetime $(M,g_{ab})$, where $M$ is the 4-dimensional spacetime manifold and
$g_{ab}$ is the spacetime metric, with signature $(-\ +\ +\ +)$. It will be assumed that the spacetime is (spatially) spherically symmetric, i.e.\ its set of isometries contains the group $SO(3)$, and all the orbits of the $SO(3)$ action are spacelike.
We also assume that the spacetime has an {\it outer trapping horizon} $\mathcal{H}$ and a {\it Kodama vector
field}. Thus, we assume that $M$ contains an open subset $N$, diffeomorphic to $\mathcal{L} \times S^2$
with an open, connected subset $\mathcal{L}$ of $\mathbb{R}^2$,
on which
advanced Eddington-Finkelstein coordinates $(v,r,\vartheta,\varphi)$ can be introduced,
where $(v,r)$ are coordinates on $\mathcal{L}$ and $(\vartheta,\varphi)$ are angular coordinates for the sphere.
The spherical symmetry group then acts on the $S^2$ part of $N$. 
Using such coordinates, the metric $g_{ab}$ on $N$ assumes the line-element
\begin{equation}\label{eq:metric}
ds^2  =   -e^{2\Psi(v,r)}C(v,r)dv^2  +   2e^{\Psi(v,r)} dv dr    + r^2 d\Omega^2
\end{equation}
where $d\Omega^2$ is the normalized spherically symmetric Riemannian metric on $S^2$.
With respect to angular coordinates $(\vartheta, \varphi)$, one has
\begin{align}
d\Omega^2= d\vartheta^2 + (\sin\vartheta)^2 d\varphi^2\,. 
\end{align}
The coordinate $v$ takes values in a real interval and $r$ in a positive real interval; the precise form of the 
intervals depends on the smooth coordinate functions $C \ge 0$ and $\Psi$. 
Furthermore, the function $C$ can be written in terms of the 
{\it Hawking mass}  
\begin{equation}\label{eq:Hawking-mass}
\mathcal{M}(v,r) = \frac{r}{2}\left(1 - g^{ab}(v,r)\nabla_a r \nabla_b r \right)
\end{equation}
as
\begin{align}
C(v,r)= 1-\frac{2 \mathcal{M}(v,r)}{r}\,.
\end{align}
As a side remark, we notice that, if $\Psi(v,r) = 0$ 
and $\mathcal{M}(v,r)= \mathcal{M}(v)$, the metric $g_{ab}$ reduces to the Vaidya metric, which is one of the simpler models of dynamical black holes (see \cite{GrifPod} and references cited there). 

Consider the following null vector fields,
\begin{align}
\ell^a = 2\frac{\partial}{\partial v}^a +e^{\Psi} C\frac{\partial}{\partial r}^a\,, \qquad 
\underline{\ell}^a = -e^{-\Psi}\frac{\partial}{\partial r}^a\,.
\end{align}
Then $\ell^a$ is a future pointing outgoing null vector field and $\underline{\ell}^a$ a future pointing ingoing null vector field. Furthermore, 
in terms of these vector fields  we have that 
\begin{align}
{g}^{ab} = - \frac{1}{2} \left( \ell^{a} \underline{\ell}^{b} + \underline{\ell}^{a}\ell^{b} \right) + h^{ab}
\end{align}
where $h_{ab} = r^2 d\Omega^2$ and it also holds that
\begin{equation}\label{eq:ell-normalization}
 g_{ab}\ell^a \underline{\ell}^b = -2.
\end{equation}

The expansion parameters of the congruences of outgoing and ingoing null geodesics tangent to $\ell^a$ and $\underline{\ell}^a$ 
are given by
\begin{gather}
\theta_+ = h^{ab}\nabla_a \ell_b = \frac{2 e^{\Psi}C}{r}  , \qquad
\theta_- = h^{ab}\nabla_a \underline{\ell}_b = - \frac{2 e^{-\Psi}}{r}.
\end{gather}
On $N$, $\theta_-$ is always negative, while $\theta_+$ has the same sign as $C$, and vanishes if $C=0$.
In the case of a black hole we have that $C$ is positive far from the center, thus
far from $r=0$, the transversal area of the congruence tangent to
$\ell^a$ increases towards the future while the transversal area of the congruence tangent to $\underline{\ell}^a$ decreases.

\subsection{Outer trapping horizons}\label{se:outer-trapping-horizon}

The set of points where $C=0$ is the union of trapped surfaces.
The possibility arises that this set has several disjoint connected components. 
Thus, we define the {\it outer trapping horizon}  $\mathcal{H}$ as 
the outermost connected component, in the following sense: We assume that there is a time function $T$ on $N$ so that 
all the $r$-coordinate values of $\mathcal{H}$ on hypersurfaces of constant $T$ are larger than the respective $r$-coordinate values of the 
other connected components.
 If there is only one connected component then, writing $C = 0$ in terms of the Hawking mass, 
\begin{equation}\label{eq:apparent-horizon}
\mathcal{H} = \left\{(v,r,\nn)\in N :\quad   \frac{2 \mathcal{M}(v,r)}{r}=1   \right\}\,.
\end{equation} 
The hypersurface $\mathcal{H}$ is spacelike for black holes which are growing in a collapse, it is lightlike for stationary black holes and it is 
timelike for black holes which evaporate.  

In contrast to the case of Schwarzschild spacetime, 
there is in general no timelike or causal Killing vector field near $\mathcal{H}$ that could be used
to define and test  black hole thermodynamical quantities.
 Hayward \cite{Hayward} proposed to use the {\it Kodama vector field} as a replacement (see also \cite{Helou} for a review). 
The Kodama vector field \cite{Kodama} can be defined in terms of $\ell$, $\underline{\ell}$ and $r$ as  
\begin{equation}\label{eq:Kodama}
 K^a :=  \frac{1}{2} \left( \ell[r]\underline{\ell}^a - \underline{\ell}[r]\ell^a  \right) = e^{-\Psi} \frac{\partial}{\partial v}^a\,,
\end{equation}
with $\ell[r] = \ell^a \nabla_a r$, and similarly for $\underline{\ell}[r]$.
The Kodama vector field is conserved and can be used to build other conserved quantities: It holds that
\begin{align}
\nabla_a K^a = 0\,, \qquad \nabla_a (W^{ab}K_b) = 0,
\end{align}
for any symmetric tensor field $W^{ab}$ that is invariant under the spherical symmetries of the spacetime.

Notice that $K^a$ is timelike in the region where $\theta_+>0$ and is lightlike on $\mathcal{H}$. 
Furthermore, the {\it surface gravity} $\kappa$ associated to the Kodama vector field is a function on $\mathcal{H}$ defined by 
\begin{equation}\label{eq:kappa}
\frac{1}{2}K^a (\nabla_a K_b-\nabla_b K_a) = \kappa K_b  \quad \text{on} \ \ \mathcal{H}\,.
\end{equation}
With respect to the metric component function $C(v,r)$ of \eqref{eq:metric}, one obtains
\begin{equation}\label{eq:kappa-eddfin}
\kappa = \frac{1}{2} \frac{\partial C(v,r)}{\partial r} \quad \text{if} \ \ C(v,r) = 0,
\end{equation}
and in terms of the Hawking mass it is
\begin{align}
\kappa = 
\left. \left( \frac{\mathcal{M}(v,r)}{r^2}
- \frac{\partial_r \mathcal{M}(v,r)}{r}
\right) \right|_{r=2\mathcal{M}(v,r)}.
\end{align}
This definition generalizes the concept of surface gravity known for stationary black hole horizons, or for bifurcate Killing
horizons. 
In the case of a stationary black hole, it is known that the surface gravity is proportional to the Hawking temperature.

\subsection{Lightlike congruences emanating from the outer trapping horizon and adapted null coordinates} \label{sec:OTH}

We have already indicated in the Introduction that there are lightlike congruences emanating from the outer trapping horizon $\mathcal{H}$ .
They are determined once one chooses any point $(v_*,r_*,\nn_*)$ on $\mathcal{H}$. Any such point then determines its
orbit $S_* = \{ (v_*,r_*) \} \times S^2$  under the spherical symmetry group
of the spacetime. Clearly, $S_*$ is a subset of $\mathcal{H}$. The lightlike vector fields $\ell^a$ and $\underline{\ell}^a$ restricted to $S_*$
are tangent to two lightlike congruences $\mathcal{C}_*$ (``outgoing'') and $\mathcal{T}_*$ (``ingoing''), respectively. Owing to the spherical symmetry,
these lightlike conguences are 2-dimensional lightlike hypersurfaces. It holds that $S_* = \mathcal{C}_* \cap \mathcal{T}_*$.

One can introduce local coordinates $(U,V)$ covering an open neighbourhood of $(v_*,r_*)$ in the $\mathcal{L}$-part of $N$.
$U$ and $V$ are null (or lightlike) coordinates, so that the metric line element \eqref{eq:metric} takes the form
\begin{equation}
 ds^2 = -2A(U,V)dU\,dV + r^2(U,V) d\Omega^2 \,,
\end{equation}
where the radial coordinate is now a function of $U$ and $V$, $r = r(U,V)$.
This can actually always be achieved for a spherically symmetric spacetime metric; in the case at hand, 
there is an integrating factor $\alpha = \alpha(v,r)$ so that the required coordinates can be defined 
on an open neighbourhood of $S_*$ by 
\begin{align}
\alpha \cdot dU_a =  \frac{1}{2} e^{2\Psi(v,r)}C(v,r)dv_a   - e^{\Psi(v,r)} dr_a   \,  , \qquad    dV_a = dv_a  \,.
\end{align}  
One can further re-define the coordinates $U$ and $V$ so that they have additional properties. First, we have the freedom
to choose the $U$ and $V$ coordinates such that $U = 0$ and $V = 0$ exactly for the points in $S_*$. Furthermore, we can choose
the $U$ and $V$ coordinates so that $U= 0$ exactly for the points on $\mathcal{C}_*$ and $V = 0$ exactly for the points on $\mathcal{T}_*$;
this freedom of choice is related to the fact that we have $\ell_a = \beta(U) \cdot dU_a$ on $\mathcal{T}_*$ and $\underline{\ell}_a = \underline{\beta}(V) \cdot
dV_a$ on $\mathcal{C}_*$, with smooth, non-zero functions $\beta(U)$ and $\underline{\beta}(V)$. Re-defining $U$ and $V$ again so that $\beta = 2$ and 
$\underline{\beta} = 1$, one obtains from \eqref{eq:ell-normalization} that
\begin{equation}\label{eq:A0}
 A(U,0) = 1 \quad \text{and} \quad A(0,V) = 1 \,.
\end{equation}
Thus, given $S_*$, we can choose coordinates $(U,V,\vartheta,\varphi) = (U,V,\nn)$ 
in an open neighbourhood of $S_*$ with the properties (1) to (4) stated in the introduction. In the next section we shall see that also property (5) is satisfied.

\subsection{Kodama flow near $S_*$}\label{se:kodamaflow}

As discussed above, once a sphere $S_*$ contained in the outer trapping horizon $\mathcal{H}$ is fixed we can determine the 
cone $\mathcal{C}_*$ formed by outgoing radial null geodesics passing through $S_*$, and the transversal cone $\mathcal{T}_*$, formed by ingoing radial null geodesics passing through $S_*$.
Later we shall analyze the scaling towards $\mathcal{C}_*$ of the 2-point function of any Hadamard state.
The resulting distribution can be restricted to  $\mathcal{T}_*$ and it will be tested with respect to an observer moving along the integral line of the Kodama field. 
Hence, we need to analyze the form of the action of $K^a$ on $\mathcal{T}_*$ near $S_*$. We denote by $\Phi_\tau$ $(\tau \in \mathbb{R})$
the flow generated by $K^a$. We recall that this means that, whenever $p \in N$ and  $\tau_0 \in \mathbb{R}$ so that 
$\Phi_{\tau}(p) \in N$ for all $\tau$ in an open interval around $\tau_0$, it holds that 
$K^a \nabla_a f |_{\Phi_{\tau_0}(p)} = \frac{d}{d\tau}|_{\tau = \tau_0} f(\Phi_{\tau}(p))$ for all smooth, real-valued functions $f$
on $N$. 

We use adapted coordinates $(U,V,\nn)$ with respect to $S_*$ as described in the previous section. 
Then we write
\begin{align}
\Phi_\tau(U,V,\nn) = ( 
                                {\rm u}_\tau(U,V,\nn),
                                {\rm v}_\tau(U,V,\nn),\nn
                               )
\end{align}
i.e.\ ${\rm u}_\tau(U,V,\nn)$ is the $U$-coordinate of $\Phi_\tau(U,V,\nn)$ and ${\rm v}_\tau(U,V,\nn)$ is the $V$-coordinate. (Note that 
$\Phi_\tau$ doesn't act on $\nn$, and therefore ${\rm u}_\tau$ and ${\rm v}_\tau$ actually do not depend on $\nn$.)
\begin{lemma} \label{lm:Kodama-flow}
 With $\kappa_*$ the value of $\kappa$ on $S_*$ (corresponding to $U = 0$ and $V = 0$), there is an open interval of $U$ coordinate
 values around 0 so that,
on using the notation $O(\tau,U^2) = O(\tau)\cdot O(U^2)$ with the usual meaning of the Landau symbol for a single argument,
 \begin{itemize}
  \item[{\rm (a)}] $dU_a K^a(U,V=0,\nn) = -\kappa_* U + O(U^2)$\,,
  \item[{\rm (b)}] ${\rm u}_\tau(U,V = 0,\nn) = {\rm e}^{- \kappa_* \tau}U + O(\tau,U^2)$ 
 \end{itemize}

\end{lemma}
\noindent
{\it Proof.} On using that $\underline{\ell}^a =\frac{\partial}{\partial U}{}^a$ and the definition of $K^a$, 
one obtains that $dU_a K^a = -\frac{1}{2}\ell[r] = \partial_\tau {\rm u}_\tau$. On the other hand, $\frac{1}{2}\ell[r]$
vanishes on $S_*$, and it holds that 
\begin{align}
\left.-\frac{1}{2}\underline{\ell}[\ell[r]]\right|_{S_*}=\left.\frac{1}{2} \frac{\partial C}{\partial r}\right|_{S_*}=\kappa_*\,.
\end{align}
This yields $dU_aK^a(U,V = 0,\nn) = -\kappa_*U + O(U^2)$, having used $\underline{\ell}^a =\frac{\partial}{\partial U}{}^a$ once more
and the fact that $S_*$ is the locus of $U= 0$ and $V= 0$. This proves (a). 

Furthermore, we have $\partial_\tau {\rm u}_\tau(U,V= 0,\nn) = dU_aK^a(U,V=0,\nn) = -\kappa_*U + O(U^2)$, yielding on integration
$$ {\rm u}_\tau(U,V = 0,\nn) = {\rm e}^{- \kappa_* \tau}U + O(\tau,U^2) $$
on fixing the constants of integration such that ${\rm u}_0(U,0,\nn) = (U,0,\nn)$ to be consistent with $\Phi_{\tau = 0}(p) = p$. This proves (b). 
${}$ \hfill $\Box$
\\[6pt]
This shows that the property (5) stated for adapted coordinates in the Introduction is also fulfilled.

\section{The quantized linear scalar field }\label{se:QF}

The main point of our article is an investigation of quantized fields on a spherically symmetric spacetime $(M,g_{ab})$ with an outer trapping horizon
$\mathcal{H}$ and Kodama vector field $K^a$. To this end, our investigation starts with the free quantized scalar field $\phi(x)$. We assume that the underlying
spacetime $(M,g_{ab})$ is globally hyperbolic. Actually, global hyperbolicity of the spacetime at large distances from $\mathcal{H}$ is 
not required for our considerations; what we need is a spherically symmetric, globally hyperbolic open 
neighbourhood of the outer trapping horizon $\mathcal{H}$ contained in the open set $N \simeq \mathcal{L} \times S^2$ on which the Eddington-Finkelstein coordinates 
discussed before can be introduced. For notational convenience, we assume in the following that this spherically symmetric globally hyperbolic open neighbourhood just coincides with $M$. 

The quantized real free scalar field on $(M,g_{ab})$ is then defined in the standard manner which we will briefly sketch. For a fuller discussion,
the reader may consult \cite{Wald2,HollandsWald,KhavkineMoretti}. As $(M,g_{ab})$ is 
globally hyperbolic by assumption, there are uniquely determined advanced and retarded fundamental solutions $G^{\rm adv/ret}$ (``Green's operators'') 
for the 2nd order hyperbolic
Klein-Gordon operator $\nabla^a\nabla_a - \mpof$ defined on smooth scalar test-functions on ${M}$. Here, $\nabla_a$ denotes the 
covariant derivative of the spacetime metric $g_{ab}$, and $\mpof \equiv \mpof(x)$ is a smooth, real-valued function on $M$.
Then one can define the causal Green's function
$$ \mathscr{G}(F,F') = \int_{M} \left( F(x) (G^{\rm adv}F')(x) - F(x)(G^{\rm ret}F')(x) \right)\, d{\rm vol}_g(x) \,, \quad F,F' \in C_0^\infty(M,\mathbb{R})\,,$$
where $d{\rm vol}_g$ denotes the volume form of the spacetime metric $g_{ab}$. Hence there is a $*$-algebra $\mathscr{A} = \mathscr{A}(M,g_{ab},\mpof)$
which is generated by a family 
of elements $\phi(F)$, $F \in C_0^\infty(M,\mathbb{R})$, and a unit element ${\bf 1}$, subject to the relations:
\begin{align*}
 (i)&  \ \ F \mapsto \phi(F) \ \ \text{is}\ \mathbb{R}\text{-linear} \quad \quad (ii) \ \ \phi(( \nabla^a\nabla_a + \mpof)F)= 0 \\
 (iii)& \ \ \phi(F)^* = \phi(F)  \quad \quad \quad \quad \quad (iv) \ \ [\phi(F),\phi(F')] = i\mathscr{G}(F,F') \cdot {\bf 1}\,, \quad \quad F,F' \in C_0^\infty(M,\mathbb{R})\,.
\end{align*}
Here, $[\phi(F),\phi(F')] = \phi(F)\phi(F') - \phi(F')\phi(F)$ denotes the algebraic commutator. The $\phi(F)$ are abstract field operators, at this level without
a Hilbert space representation. One can symbolically write $\phi(x)$ to mean that $\phi(F) = \int_{M} \phi(x)F(x)\,d{\rm vol}_g(x)$ which can best be made
rigorous when the $\phi(F)$ are given in some Hilbert space representation. 
\par
We recall that $w^{(2)}$ is a 2-point function for the Klein-Gordon field operators $\phi(F)$ if $w^{(2)} : C_0^\infty(M,\mathbb{R}) \times C_0^\infty(M,\mathbb{R}) \to \mathbb{C}$, 
$F,F' \mapsto w^{(2)}(F,F')$ is real-bilinear, extends to a distribution in $\mathcal{D}'(M \times M)$, and moreover fulfills
\begin{align}
& w^{(2)}(F,F) \ge 0\,, \quad w^{(2)}(F',F) = \overline{w^{(2)}(F,F')}\,, \quad {\rm Im}\,w^{(2)}(F,F') = {\frac{1}{2}}\mathscr{G}(F,F')\,, \\
  & w^{(2)}((\nabla^a\nabla_a - \mpof)F,F') = 0 = w^{(2)}(F,(\nabla^a\nabla_a - \mpof)F') \quad \quad (F,F' \in C_0^\infty(M,\mathbb{R})\,.
\end{align}

There is a one-to-one correspondence between states on $\mathscr{A}$ and Hilbert space representations of $\mathscr{A}$ which
is given by the Gelfand-Naimark-Segal (GNS) representation. At this point, however, we don't make use of this, but we come back to 
a more operator-algebraic point of view in Sec.\ \ref{se:EntAr-SL}. Rather, we are interested 
in quasifree Hadamard states on $\mathscr{A}$; as these are completely determined by their 2-point function $w^{(2)}$, it is the 
behaviour of these 2-point functions near the outer trapping horizon that will be in the focus of our investigation.
\par
At this point it is useful, for later purpose, to look at the Hadamard form of the 2-point function in more detail,
following mainly \cite{KayWald,Rad1,SahlmannVerch} (see also \cite{SahlmannVerch0} for the relation of the Hadamard condition with equilibrium states).
\\[6pt]
Having chosen some $S_*$, in the adapted coordinates we can define the time-function $T(x) = T(U,V,\vartheta,\varphi) = (U + V)/2$. We can consider the 
hypersurface $\Sigma = \{ x = (U,V,\vartheta,\varphi) : T(x) = 0\}$. Then there is an open neighbourhood $\mathcal{B}$ in $M$ of $S_*$ so that 
$\Sigma_{\mathcal{B}} = \Sigma \cap \mathcal{B}$ is a spacelike, acausal hypersurface containing $S_*$, and 
the open interior of the domain of dependence
$D(\Sigma_{\mathcal{B}})$ is 
a globally hyperbolic open neighbourhood of $S_*$ having $\Sigma_{\mathcal{B}}$ as a Cauchy-surface. Then an open neighbourhood 
$N_{\mathcal{B}}$ of 
$\Sigma_{\mathcal{B}}$ is called a {\it causal normal neighbourhood} if, given $x$ and $x'$ in $N_{\mathcal{B}}$, with $x' \in J^+(x)$, there is a convex normal neighbourhood (with respect to the 
metric $g_{ab}$) containing $J^-(x') \cap J^+(x)$. It has been shown in \cite{KayWald} that causal normal neighbourhoods always exist. 
\\[6pt]
Then a 2-point function $w^{(2)}$ is said to be {\it of Hadamard form near $S_*$} if, for all $F,F' \in N_{\mathcal{B}}$, it holds that 
\begin{align}
 w^{(2)}(F,F') = \lim_{\varepsilon \to 0+} \int w_\epsilon(x,x') F(x)F'(x')
\,d{\rm vol}_g(x)\,d{\rm vol}_g(x') \quad \ \ (F,F' \in C_0^\infty(N_{\mathcal{B}},\mathbb{R})) \,,
\end{align}
where (for any $k \in \mathbb{N}$)
\begin{align} \label{Hdaform}
  w_\varepsilon(x,x') = \psi(x,x')\left(  \frac{1}{8\pi^2}\frac{{\Delta}^{1/2}(x,x')}{\sigma_\varepsilon(x,x')} +  \ln(\sigma_\varepsilon(x,x'))Y_k(x,x') \right) + Z_k(x,x')
\end{align}
with 
\begin{align}
 \sigma_\varepsilon(x,x') = \sigma(x,x') - 2i \varepsilon t(x,x') + \varepsilon^2 \quad \text{and} \quad t(x,x') = T(x) - T(x').
\end{align}
The important point here is the appearance of a smooth cut-off function $\psi$ whose purpose
is to make the terms in the brackets in \eqref{Hdaform} well-defined and smooth. We explain this here briefly and refer for 
full details to the references \cite{KayWald,Rad1,SahlmannVerch} and
also the review \cite{KhavkineMoretti}; note that in the references, our cut-off function
$\psi$ is denoted by $\chi$ (however, we use $\chi$ as a different cut-off function in the proof
of Thm.\ \ref{Thm:SL}).\footnote{We thank an anonymous referee for pointing out that 
a previous version of our definition of the cut-off function was incomplete.}
We denote by $\mathcal{X}$ the set of causally related pairs of points $(x,x') \in N_{\mathcal{B}} \times N_{\mathcal{B}}$, i.e.\
$(J^+(x) \cap J^-(x')) \cup (J^-(x) \cap J^+(x'))$ is non-empty. Then there is an open neighbourhood $\mathcal{U}$ of $\mathcal{X}$ in
$N_{\mathcal{B}} \times N_{\mathcal{B}}$ on which $\sigma(x,x')$ (the half of the squared geodesic distance) and $\Delta(x,x')$ (the 
van Vleck-Morette determinant) are well-defined and $C^\infty$ for $(x,x') \in \mathcal{U}$ --- however, it seems that a proof 
of existence of such a neighbourhood $\mathcal{U}$ has never previously been given in the literature. That issue is discussed 
in a recent paper by Moretti \cite{Mor21}, where an argument showing that actually there is such a $\mathcal{U}$ 
is presented. 
Furthermore, there is a
well-defined sequence of functions $Y_k(x,x')$ (determined by the Hadamard recursion relations) which can be chosen in $C^k(\mathcal{U},\mathbb{R})$ for any $k \in \mathbb{N}$. The functions $Z_k(x,x')$ are correspondingly in $C^k(N_\mathcal{B} \times N_\mathcal{B},\mathbb{R})$.
There is then a further open neighbourhood of $\mathcal{X}$ in $N_\mathcal{B} \times N_\mathcal{B}$, denoted by $\mathcal{U}_*$, so that 
$\overline{\mathcal{U}_*} \subset \mathcal{U}$. Then choose some $\psi \in  C^\infty(N_{\mathcal{B}} \times N_{\mathcal{B}},[0,1])$ with
$\psi(x,x') = 1$ on $\mathcal{U}_*$ and $\psi(x,x') = 0$ outside of $\mathcal{U}$. Consequently, the bracket term in \eqref{Hdaform} is 
well-defined and $C^k$ due to the presence of the cut-off function $\psi$. There is a freedom of choice for $\psi$; a different choice 
is compensated by a re-definition of the $Z_k$. Otherwise, different sequences $Z_k$ correspond to different two-point functions.

\section{Scaling limit of Hadamard 2-point functions near $S_*$ and restriction to $\mathcal{T}_*$}

\subsection{Conformal transformation}

In the adapted coordinates $(U,V,\vartheta,\varphi)$ discussed in Sec \ref{sec:OTH}, the line element of the spacetime $(M,g_{ab})$ under
consideration assumes the form 
$$ 
ds^2 = -2A(U,V)dUdV + r^2(U,V) d\Omega^2 \,.
$$
Our investigation of the scaling limit of the quantized linear scalar field near points of an outer trapping horizon $\mathcal{H}$ that we will consider below 
will be facilitated by using a conformally transformed metric. This applies in particular to the proof of Thm.\ \ref{Thm:SL}. 
To simplify notation, we will write again $\nn$ for $  (\vartheta,\varphi)$
noting that the angular variables $(\vartheta,\varphi)$ really represent an element $\nn$ of the unit sphere. 

The conformal transformation is defined with respect to an arbitrarily chosen point $(v_*,r_*)$ on $\mathcal{H}$
defining $S_*$ and consequently $\mathcal{C}_*$ and $\mathcal{T}_*$, as explained in the Introduction. Given $(v_*,r_*)$ (or equivalently, the 
corresponding $S_*$), we introduce the 
conformally transformed metric $\tilde{g}_{ab}$ on $M$ by 
\begin{align} \label{eq:conf-metric}
 \tilde{g}_{ab} = \eta^2 g_{ab} \ \ \quad 
\text{with the conformal factor} 
\ \  \quad 
\eta^2(U,V) = \frac{r_*^2}{r^2(U,V)} \,\text{;}
\end{align}
the associated line element is
\begin{align} \label{eq:conformal-metric}
 d\tilde{s}{}^2 = -2 \frac{A(U,V) r_*^2}{r^2(U,V)} dU dV + r_*^2 d\Omega^2 \,.
\end{align}
One feature of $\tilde{g}_{ab}$ is the splitting of the squared geodesic distance between points $(U,V,\nn)$ and 
$(U',V',\nn')$ according to the Pythagorean theorem:
\begin{equation}\label{eq:pitagora}
 \tilde{\sigma}(U,V,\nn;U',V',\nn') = \tilde{\sigma}{}^{(\mathcal{L})}(U,V;U',V') + {\sf s}(\nn;\nn') 
\end{equation}
where $\tilde{\sigma}{}^{(\mathcal{L})}(U,V;U',V')$ denotes the squared geodesic distance between the points  $(U,V)$ and $(U',V')$ on the two-dimensional ``Lorentzian'' part of 
the spacetime with metric line element $-2 \frac{A(U,V) r_*^2}{r^2(U,V)} dU dV$, and where ${\sf s}(\nn;\nn')$ is the 
squared geodesic distance between points $\nn$ and $\nn'$ on the two-dimensional sphere with radius $r_*$.
\par
It is worth noting that on $S_*$ where $r = r_*$, the conformal factor is equal to 1:
$ \left. \eta \right|_{S_*} = 1 \,. $
\par
At the level of 2-point functions, the conformal transformation has the following effect.
Suppose that 
$$ 
w^{(2)}(F,F') = \lim_{\varepsilon \to 0}\, \int w_\varepsilon(x,x') F(x) F'(x')\, d{\rm vol}_g(x)\,d{\rm vol}_g(x') \quad \ \
(F,F' \in C_0^\infty(N_{\mathcal{B}},\mathbb{R}))
 $$
is a 2-point function of Hadamard form, near $S_*$, for the quantized scalar field that we consider on $(M,g_{ab})$. Then, defining 
$$
\tilde{w}_\varepsilon(x,x') = \eta^{-1}(x) w_\varepsilon(x,x') \eta^{-1}(x') \,,
$$
the distribution 
$$
\tilde{w}{}^{(2)}(F,F') = \lim_{\varepsilon \to 0}\, \int \tilde{w}_\epsilon(x,x') F(x) F'(x')\, d{\rm vol}_{\tilde{g}}(x)\,d{\rm vol}_{\tilde{g}}(x') \quad \ \
(F,F' \in C_0^\infty(N_{\tilde{\mathcal{B}}},\mathbb{R})),
 $$
is a 2-point function of Hadamard form near $S_*$ on the conformally related spacetime $(M,\tilde{g}_{ab})$, with 
a suitably small neighbourhood $\tilde{\mathcal{B}}$ of $S_*$, and an associated causal normal neighbourhood $N_{\tilde{\mathcal{B}}}$ 
defined with respect to $\tilde{g}_{ab}$. This has been shown in \cite{P}.

The scaling limit which we will consider in the next section gives the same results on $w^{(2)}$ or on $\tilde{w}^{(2)}$ on
account of $\left. \eta \right|_{S_*} = 1$, but it is easier to study the scaling limit using $\tilde{w}^{(2)}$ because of 
\eqref{eq:pitagora}. To this end, we 
put on record the following observations for later use. 
\\[6pt]
The volume form $d{\rm vol}_{g}$ of the original metric and the volume form $d{\rm vol}_{\tilde{g}}$ of the conformally transformed 
metric are related according to
$
 d{\rm vol}_{\tilde{g}}(x) = \eta^4(x) d{\rm vol}_g(x)
$
and therefore one has 
\begin{align} \label{eq:match}
 w^{(2)}(F,F') = \tilde{w}{}^{(2)}(\tilde{F},\tilde{F}') \quad \ \ \text{with} \ \, \tilde{F} = \eta^{-3} F\,, \ \tilde{F}' = \eta^{-3}F
\end{align}
for all $F,F' \in C_0^\infty(M,\mathbb{R})$. The statement that $\tilde{w}{}^{(2)}$ is of Hadamard form near $S_*$ on $(M,\tilde{g}_{ab})$ means that 
\begin{align} \label{eq:conf-Hadform}
\tilde{w}_\varepsilon(x,x') = \tilde{\psi}(x,x') \left( \frac{1}{8\pi^2}\frac{\tilde{\Delta}{}^{1/2}(x,x')}{\tilde{\sigma}_\varepsilon(x,x')} + 
\ln(\tilde{\sigma}_\varepsilon(x,x'))\tilde{Y}(x,x') \right) + \tilde{Z}(x,x') \quad (x,x' \in N_{\tilde{\mathcal{B}}})
\end{align}
where $\tilde{\Delta}$ and $\tilde{\sigma}$ refer to $\tilde{g}_{ab}$, $\tilde{\psi}$ has properties analogous to $\psi$, and $\tilde{Y}$ and $\tilde{Z}$ 
(dropping the index $k$ on $Y$ and $Z$)
can be 
chosen as $C^k$ function for any $k \in \mathbb{N}$.

\subsection{Scaling limit and restriction}\label{se:scaling}
We select a sphere $S_*$ of radius $r_*$ lying in the outer trapping horizon, and a patch of adapted coordinates $(U,V,\nn)$ relative to 
$S_*$. Moreover, we assume that $N_{\mathcal{B}}$ is a causal normal neighbourhood of a partial Cauchy surface $\Sigma_{\mathcal{B}}$
so that $S_* \subset \Sigma_{\mathcal{B}}$, as described in Sec. \textcolor{red}{\ref{se:QF}}.
Then if $l_0 > 0$ is small enough, the open set 
\begin{align}
 \mathcal{O} = \{(U,V,\nn) : |U| < l_0\,, \ \, |V| < l_0\,, \ \, \nn \in S^2\} 
\end{align}
is a subset of $N_{\mathcal{B}}$ and an open neighbourhood of $S_*$.
We assume that $l_0$ is chosen small enough so that $\mathcal{O}$ is also contained in a causal normal neighbourhood $N_{\tilde{\mathcal{B}}}$ of 
$S_*$ with respect to the
conformally transformed metric $\tilde{g}_{ab}$ described in Sec.\ 4.1.

Note that, if $0 < \lambda \le 1$ and $0 < \mu \le 1$ then
 $(U,V,\nn) \in \mathcal{O} \Rightarrow  (\lambda U, \mu V, \nn) \in \mathcal{O}$.
Consequently, when defining the {\it scaling transformations} 
\begin{align}
 ({\sf u}_\lambda F)(U,V,\nn) = \frac{1}{\lambda}F(U/\lambda,V,\nn)\,, 
\end{align}
for $0 < \lambda \le 1$, one can see that the ${\sf u}_\lambda$  map the space of test functions
$C_0^\infty(\mathcal{O},\mathbb{R})$ into itself, and
\begin{align}
{\rm supp}({\sf u}_\lambda F) = \{ (\lambda U, V,\nn) : (U,V,\nn) \in {\rm supp}(F)\} \quad
(0 < \lambda \le 1)\,.
\end{align}
We stress that the scaling transformations are defined with respect to the chosen $S_*$, and the corresponding adapted coordinates.

We also define another type of transformations which serve, in a limit, to restricting distributions to $\mathcal{T}_*$ by
effectively acting like a $\delta$-distribution concentrated at $V = 0$. Let $\zeta \in C_0^\infty((-l_0,l_0),\mathbb{R})$ with 
$\zeta(V) \ge 0$ and $\int \zeta(V)\,dV = 1$. Then we define, for any $F \in C_0^\infty(\mathcal{O},\mathbb{R})$, 
\begin{align}
 ({\sf v}_\mu F)(U,V,\nn) = \frac{1}{\mu} \zeta (V/\mu) F(U,V,\nn) \quad \ \ (0 < \mu \le 1)\,.
\end{align}
Clearly, also every ${\sf v}_\mu$ maps $C_0^\infty(\mathcal{O},\mathbb{R})$ into itself. 
\\[6pt]
Adopting this notation, we now present the result on scaling limits of Hadamard 2-point functions near $S_*$ and subsequent restriction to $\mathcal{T}_*$. 
\begin{theorem} \label{Thm:SL}
 Let $w^{(2)}$ be any 2-point function of Hadamard form for the scalar field $\phi$ on the spacetime $(M,g_{ab})$
 as in Sec.\ \ref{se:QF}.
 \\[2pt]
 {\rm (I)} \ \ For all $f,f' \in C_0^\infty(\mathcal{O},\mathbb{R})$ it holds that 
 \begin{align}
  \lim_{\lambda \to 0}&\,w^{(2)}({\sf u}_\lambda(2\partial_U f),{\sf u}_{\lambda}(2\partial_{U'}f'))  = L(f,f')\,, \quad \ \ \ \text{where}
  \\[6pt]
  L(f,f') & =
\lim_{\varepsilon\to 0+} -\frac{1}{r_*^2\pi} \int  \frac{ f(U,V,\nn) f'(U',V',\nn)}{(U-U'+i\varepsilon)^2}Q(V,V',\nn)\,
dU\,dU\,'dV\,dV'\,d\Omega^2(\nn) \nonumber
  \end{align}
with
\begin{align}
 Q(V,V',\nn) & = \tilde{\Delta}^{1/2}(0,V,\nn,0,V',\nn) P(0,V,0,V') \, ,\\
 P(U,V,U',V') &  = r_*^4\eta^{-1}(U,V)A(U,V)\eta^{-1}(U',V')A(U',V') \,.
\end{align}
{\rm (II)} \ \ 
\ \ For all $f,f' \in C_0^\infty(\mathcal{O},\mathbb{R})$ it holds that 
\begin{align}
\lim_{\mu \to 0}  \lim_{\lambda \to 0}\,w^{(2)}({\sf v}_\mu{\sf u}_\lambda({2}\partial_U f),{\sf v}_\mu{\sf u}_{\lambda}({2}\partial_{U'}f'))
= \lim_{\mu \to 0} L({\sf v}_\mu f,{\sf v}_\mu f') = \Lambda(f,f')
\end{align}
where
\begin{align}\label{eq:Lambda}
\Lambda(f,f') =
\lim_{\varepsilon\to 0+} -\frac{r_*^2}{{\pi}} \int  \frac{ f(U,0,\nn) f'(U',0,\nn)}{(U-U'+i\varepsilon)^2}
dU\,dU'\,d\Omega^2(\nn)
  \end{align}
\end{theorem}
\noindent
The proof of this Theorem will be given the Appendix (Sec.\ \ref{Appendix}).
\\[6pt]
{\bf Remark}
\\[2pt]
{\sf (i)} \ \ The more difficult step is proving Part (I) of the Theorem, Part (II) then is merely a corollary. 
Actually, the statement follows easily when inserting the 
scaled test functions ${\sf u}_\lambda({2}\partial_U f)$ and ${\sf u}_\lambda({2}\partial_{U'} f')$ into the $\varepsilon$-regulated
integral expression of the Hadamard form and exchanging the $\lambda \to 0$ and $\varepsilon \to 0$ limits. The more 
involved part of the proof consists in showing that this can be justified. We have opted to give a full, self-consistent proof 
in this article, despite some similarities of our proof with a related argument in \cite{MP} (that relied in parts also on results from \cite{KayWald}) which applies 
to the case of the quantized Klein-Gordon field on spacetimes with bifurcate Killing horizons. 
\\[2pt]
{\sf (ii)} \ \ As is familiar from the quantization of the massless free quantum field in 2-dimensional Minkowski spacetime, respectively its chiral 
components on lightrays, the 2-point function is well-defined for test-functions which are first derivatives of compactly supported smooth
functions. Without derivatives, an infrared divergence occurs, see e.g.\ \cite{BLTO}, Sec.\ on the ``Schwinger model''. This is the reason why 
the test-functions used for the scaling limit considerations are $U$-derivatives of compactly supported smooth functions. 
\\[2pt]
{\sf (iii)} \ \ One may choose $U$ or $V$-coordinates so that the Van Vleck\,-\,Morette determinant is equal to 1;
this simplifies the form of the function $Q$ in the first part of Theorem \ref{Thm:SL}, however we need not make use 
of this possibility here. 
\\[2pt]
{\sf (iv)} \ \ The factor 2 in the definition of the scaling transformations ${\sf u}_\lambda$ has been introduced to match $\Lambda$ with the convention for 
2-point functions on lightlike hyperplanes used 
in the literature, see e.g. \cite{DMP17}. See also the remark towards the end of Sec.\ \ref{se:rel-entropy-area}.

\subsection{Kodama flow projected to $\mathcal{T}_*$ and its action in the scaling limit}\label{se:Kodama-Flow-Projected}

Under the same assumptions as for the previous theorem, we can establish that the {\it projected action ${\rm T}_\tau$ on} $\mathcal{T}_*$ of the 
flow of the Kodama vector field $K^a$ acts like the dilation group in the scaling limit. To make this more precise, we define
\begin{align}
 {\rm T}_\tau(U,V,\nn) = ({\rm u}_\tau(U,0,\nn),V,\nn)   \quad \ \ (\tau \in \mathbb{R})
\end{align}
for all $(U,V,\nn) \in \mathcal{O}$, with the convention that the definition applies whenever $({\rm u}_\tau(U,0,\nn),V,\nn)$ is
again in $\mathcal{O}$. Recall that (cf.\ Lemma \ref{lm:Kodama-flow}) 
$ {\rm u}_\tau(U,V = 0,\nn) =  {\rm e}^{- \kappa_* \tau}U + O(\tau,U^2)$
so that the projected action of the Kodama flow on $\mathcal{T}_*$ takes the form
\begin{align}
{\rm T}_\tau(U,V,\nn) =  ({\rm e}^{- \kappa_* \tau}U + O(\tau,U^2),V,\nn)
\end{align}
and there is some $\tau_0 > 0$ and an open neighbourhood $\mathcal{O}_0$ of $S_*$ with $\mathcal{O}_0 \subset \mathcal{O}$ so that 
$T_\tau(U,V,\nn) \in \mathcal{O}$ for all $(U,V,\nn)$ in $\mathcal{O}_0$ and all $|\tau| < \tau_0$. 
We also define: 
\begin{align}
 ({\sf T}_\tau F)(U,V,\nn) & = F({\rm T}_{-\tau}(U,V,\nn)) \quad \ \ (F \in C_0^\infty(\mathcal{O}_0,\mathbb{R})\,, \ |\tau|<\tau_0)\,, \\
 {\rm S}_{\tau}(U,\nn) & = ({\rm e}^{ - \kappa_*\tau} U,\nn) \,, \quad \ \ ((U,\nn) \in \mathcal{T}_*\,, \ \tau \in \mathbb{R})\,, \\
 ({\sf S}_{\tau}\varphi)(U,\nn) & = \varphi({\rm S}_{-\tau}(U,\nn)) \quad \ \
 (\varphi \in C_0^\infty(\mathcal{T}_*,\mathbb{R})\,, \ \tau \in \mathbb{R}) 
\end{align}

\begin{proposition}\label{prop:scaled-kodama-flow}
 For any 2-point function $w^{(2)}$ of $\phi$ that is of Hadamard form,
\begin{align}
 \lim_{\mu \to 0+} \lim_{\lambda \to 0+} w^{(2)}({\sf T}_{\tau} {\sf v}_\mu {\sf u}_\lambda (\partial_U f),
 {\sf T}_{\tau'} {\sf v}_\mu {\sf u}_\lambda (\partial_{U'} f'))
  = \Lambda({\sf S}_\tau f,{\sf S}_{\tau'}f') 
\end{align}
holds for all $f,f' \in C_0^\infty(\mathcal{O}_0,\mathbb{R})$ and $|\tau|,|\tau'| < \tau_0$. 
\end{proposition}
\noindent
{\it Proof.} For any $F \in C_0^\infty(\mathcal{O}_0,\mathbb{R})$ and $|\tau| < \tau_0$, one obtains
\begin{align}
 ({\sf T}_\tau {\sf u}_\lambda F)(U,V,\nn) & = ({\sf u}_\lambda F)({\rm T}_{-\tau}(U,V,\nn)) \nonumber \\
   &  = ({\sf u}_\lambda F)( {\rm e}^{\kappa_* \tau}U + O(\tau,U^2),V,\nn) \nonumber \\
   & = \frac{1}{\lambda} F(\lambda^{-1}({\rm e}^{\kappa_* \tau} U + O(\tau,U^2)),V,\nn) \nonumber\\
   & = \frac{1}{\lambda} F({\rm e}^{\kappa_* \tau}(U/\lambda) + O(\lambda) \cdot O(\tau,(U/\lambda)^2),V,\nn)
\end{align}
for small enough $\lambda > 0$. One can now see that in the proof of Thm.\ \ref{Thm:SL}, all estimates involving 
$F_\lambda(x) = ({\sf u}_\lambda F)(U,V,\nn)$ (and similarly, the primed counterparts) are preserved when replacing 
$({\sf u}_\lambda F)(U,V,\nn)$ by $({\sf T}_\tau {\sf u}_\lambda F)(U,V,\nn)$ (and similarly for the primed counterparts).
Moreover, the limit considerations in the proof of Thm.\ \ref{Thm:SL} where $F(x) = F(U,V,\nn)$ appears (and the primed 
counterpart) render the analogous results upon replacing $F(U,V,\nn)$ by $F( {\rm e}^{ \kappa_* \tau} U + O(\lambda)\cdot 
O(\tau,U^2) , V, \nn)$ (analogously for the primed counterpart) as $\lambda \to 0$, except that $F(U,V,\nn)$ is in the 
limit to be replaced by $F({\rm e}^{\kappa_* \tau}U,V,\nn)$ and $F'(U',V',\nn')$ by $F'({\rm e}^{\kappa_* \tau'}U',V',\nn')$.
That follows from the fact that $O(\lambda)\cdot O(\tau,U^2) \to 0$ as $\lambda \to 0$ uniformly as $\tau$ and $U$ vary 
over compact sets. Observing this and carrying out the steps of the proof of Thm.\ \ref{Thm:SL} thus yields the claimed 
result. \hfill $\Box$

\section{Thermal properties and entropy-area relation for the scaling-limit-theory on $\mathcal{T}_*$} \label{se:EntAr-SL}

\subsection{The scaling-limit-theory on $\mathcal{T}_*$ (and its extension)} \label{se:SLtheory}

The 2-point function $\Lambda$ defines a quantum field theory on $\mathcal{T}_*$ which naturally extends to a 
(chiral, conformal) quantum field theory on $\mathbb{R} \times S_* \simeq \mathbb{R} \times S^2$. We will refer to this as 
the ``scaling-limit-theory'' induced by the scaling limit 2-point function $\Lambda$.

To discuss this, fix again $S_* \subset \mathcal{H}$, which is a copy of the sphere $S^2$ with radius $r_*$. Then 
one can introduce on the (real-linear) function space $\mathcal{D}_{S_*} = C_0^\infty(\mathbb{R} \times S_*,\mathbb{R})$ the symplectic form
\begin{align}
 \varsigma(\varphi,\varphi') = { 2}{\rm Im}\, \Lambda(\varphi,\varphi') = r_*^2 \int ( \partial_U \varphi(U,\nn)\varphi'(U,\nn) 
 -\varphi(U,\nn) \partial_U \varphi'(U,\nn)
 )\, dU \,d\Omega^2(\nn) \,.
\end{align}
Note the dependence of $\varsigma$ on $r_*^2$. Given this symplectic form, one can form the Weyl-algebra $\mathcal{W}(\mathcal{D}_{S_*},\varsigma)$ (the ``exponentiated 
CCR algebra'') over the symplectic space $(\mathcal{D}_{S_*},\varsigma)$; by definition, it is a $C^*$ algebra 
with unit element ${\bf 1}$, generated by 
unitary elements $W(\varphi)$, $\varphi \in \mathcal{D}_{S_*}$, fulfilling the Weyl-relations 
\begin{align}
 W(0) = {\bf 1}\,, \quad W(\varphi)^* = W(-\varphi)\,, \quad W(\varphi)W(\varphi') = {\rm e}^{{ -\frac{i}{2}}  \varsigma(\varphi,\varphi')} W(\varphi + \varphi') \,.
\end{align}
As is common in the operator algebraic approach to algebraic quantum field theory (cf.\ \cite{Haag} and in the present context, see 
also \cite{DMP17,KayWald,GLRV,SummersVerch})
one can introduce a family $\{ \mathcal{W}(G) \}$ of $C^*$ algebras indexed by open, relatively compact subsets $G$ of $\mathbb{R} \times S_*$
by defining $\mathcal{W}(G)$ as the $C^*$-subalgebra generated by all $W(\varphi)$ with ${\rm supp}(\varphi) \subset G$. Then it is 
easy to see that $\{ \mathcal{W}(G) \}$ fulfills the condition of {\it isotony}, meaning that $\mathcal{W}(G_1) \subset 
\mathcal{W}(G_2)$ if $G_1 \subset G_2$, and it fulfills also a condition of {\it locality}, which in the present case 
means that $\mathcal{W}(G_1)$ and $\mathcal{W}(G_2)$ commute elementwise if $G_1 \cap G_2 = \emptyset$.
Furthermore, there are certain symmetries that act covariantly on the manifold $\mathbb{R} \times S_*$:
The dilations ${\rm S}_\tau (U,\nn) = ({\rm e}^{-\kappa_* \tau} U,\nn)$, the translations ${\rm L}_a(U,\nn) = 
(U + a,\nn)$, and rotations ${\rm R}(U,\nn) = (U,R \nn)$, where $\tau,a \in \mathbb{R}$ and $R \in SO(3)$. The actions 
of these symmetry operations can be lifted to $\mathcal{D}_{S_*}$ by setting 
${\sf S}_\tau \varphi = \varphi\circ {\rm S}_{\tau}^{-1}$, ${\sf L}_a \varphi = \varphi \circ {\rm L}_a^{-1}$ and ${\sf R} \varphi = \varphi \circ {\rm R}^{-1}$. 
Each of those is a symplectomorphism with respect to the symplectic form $\varsigma$, i.e.\ one has 
$\varsigma({\sf S}_\tau \varphi, {\sf S}_\tau \varphi') = \varsigma(\varphi,\varphi')$ for all $\varphi,\varphi' \in \mathcal{D}_{S_*}$, etc. This implies that 
these symplectomorphisms can be lifted to $C^*$-algebraic morphisms $\alpha_{(\tau,a,R)}$ of $\mathcal{W}(\mathcal{D}_{S_*},\varsigma)$, given by 
\begin{align}
 \alpha_{(\tau,a,R)}W(\varphi) = W({\sf S}_\tau {\sf L}_a {\sf R} \varphi) \,.
\end{align}
We also adopt the notation to write $\alpha_\tau$ for $\alpha_{(\tau,0,1)}$ and $\alpha_a$ for $\alpha_{(0,a,1)}$ etc whenever 
no ambiguity can arise. It is plain that thereby, a represention of the group of symmetries generated by dilations, translations and rotations by automorphisms 
of $\mathcal{W}(\mathcal{D}_{S_*},\varsigma)$ is established. It is also easily seen that these 
automorphisms act covariantly (or geometrically) on the family $\{ \mathcal{W}(G)\}$ in the sense that 
\begin{align}
 \alpha_\tau(\mathcal{W}(G)) = \mathcal{W}({\rm S}_\tau G)\,, \quad \alpha_a(\mathcal{W}(G)) = \mathcal{W}({\rm L}_a G)\,,
 \quad \alpha_R (\mathcal{W}(G)) = \mathcal{W}({\rm R} G)\,.
\end{align}
We recall that a linear functional $\omega : \mathcal{W}(\mathcal{D}_{S_*},\varsigma) \to \mathbb{C}$ is 
a {\it state} if it is positive, i.e.\ $\omega(A^*A) \ge 0$ for all $A \in \mathcal{W}(\mathcal{D}_{S_*},\varsigma)$, and 
normalized, i.e.\ $\omega({\bf 1}) = 1$. Moreover, every state $\omega$ induces the associated GNS-representation
$(\mathscr{H}_\omega,\pi_\omega,\Omega_\omega)$ of $\mathcal{W}(\mathcal{D}_{S_*})$, characterized by the properties that 
$\pi_\omega$ is a unital $*$-representation of $\mathcal{W}(\mathcal{D}_{S_*},\varsigma)$ by bounded linear operators on the 
Hilbert space $\mathscr{H}_\omega$, and $\Omega_\omega$ is a unit vector in $\mathscr{H}_\omega$ so that 
$\pi_\omega(\mathcal{W}(\mathcal{D}_{S_*},\varsigma))\Omega_\omega$ is dense in $\mathscr{H}_\omega$ and 
$ \langle \Omega_\omega,\pi_\omega(A) \Omega_\omega\rangle = \omega(A)$ for all $A \in \mathcal{W}(\mathcal{D}_{S_*},\varsigma)$
(on writing $\langle \xi,\psi \rangle$ for the scalar product of $\xi,\psi \in \mathscr{H}_\omega$). 
Thus, once given a state $\omega$ on $\mathcal{W}(\mathcal{D}_{S_*},\varsigma)$, one can introduce the system $\{ \mathcal{N}(G) \}$
of {\it local von Neumann algebras} in the GNS-representation of $\omega$ given by 
\begin{align}
 \mathcal{N}(G) = \overline{\pi_\omega(\mathcal{W}(G))} = \pi_\omega(\mathcal{W}(G))''
\end{align}
where the overlining means taking the weak closure in $\mathcal{B}(\mathscr{H}_\omega)$ (the set of bounded linear operators on
$\mathscr{H}_\omega$); the double prime denotes the double commutant: For $\mathcal{X} \subset \mathcal{B}(\mathscr{H}_\omega)$,
$\mathcal{X}' = \{ B \in \mathcal{B}(\mathscr{H}_\omega) : AB - BA = 0 \ \ \text{for all} \ A \in \mathcal{X} \,\}$ is the 
{\it commutant} of $\mathcal{X}$, and $\mathcal{X}'' = (\mathcal{X}')'$. Whenever $\mathcal{X}$ contains the unit operator, 
it holds that $\overline{\mathcal{X}} = \mathcal{X}''$. For full details on these operator algebraic facts, see \cite{BraRo1,BraRo2,Borchers}. 
\\[6pt]
The 2-point function $\Lambda$ induces a quasifree state $\omega_\Lambda$ on $\mathcal{W}(\mathcal{D}_{S_*},\varsigma)$ 
defined by linear extension of the assignment $\omega_\Lambda(W(\varphi)) = {\rm e}^{-\Lambda(\varphi,\varphi)/2}$. We 
denote the associated local von Neumann algebras again by $\mathcal{N}(G)$ (unless a more detailed notation is required). 
Of particular interest are the von Neumann algebras $\mathcal{N}_R = \mathcal{N}( (-\infty,0) \times S_*)$ and 
$\mathcal{N}_L = \mathcal{N}((0,\infty) \times S_*)$. 

Several important properties of $\omega_\Lambda$ have been established and are well-known, from related contexts or from
investigations of chiral conformal quantum field theory. We collect some of those properties here; proofs and further exposition
can be found in \cite{DMP17,KayWald,GLRV,SummersVerch}. For notational simplicity, the GNS representation of $\omega_\Lambda$
will be denoted by $(\mathscr{H}_\Lambda,\pi_\Lambda,\Omega_\Lambda)$. 
\begin{itemize}
 \item[(1)] The state $\omega_\Lambda$ is invariant under the action $\alpha$: $\omega_\Lambda \circ \alpha_{(\tau,a,R)} = \omega_\Lambda$\,.
 Consequently, there is a unitary action ${\sf U}_{(\tau,a,R)}\pi_{\Lambda}(A)\Omega_{\Lambda} = 
 \pi_{\Lambda}(\alpha_{(\tau,a,R)}A)\Omega_{\Lambda}$ $(A \in \mathcal{W}(\mathcal{D}_{S_*},\varsigma))$ implementing 
 the action of $\alpha$ in the GNS-representation of $\omega_\Lambda$ with ${\sf U}_{(\tau,a,R)}\Omega_{\Lambda}
 = \Omega_{\Lambda}$.
 \item[(2)] $\omega_\Lambda$ is a ground state for the translations $\alpha_a$, i.e.\ there is a 
 non-negative selfadjoint generator ${\sf H}$ in $\mathscr{H}_{\Lambda}$ so that ${\sf U}_a = {\rm e}^{i {\sf H}a}$.
 (We are here using the same convention as previously explained for $\alpha$ to write ${\sf U}_a = {\sf U}_{(0,a,1)}$, etc.)
 \item[(3)] $\Omega_\Lambda$ is a cyclic and separating vector for the von Neumann algebras $\mathcal{N}_R$ and 
 $\mathcal{N}_L$. Let $\Delta_R$ denote the modular operator with respect to $\mathcal{N}_R$ and 
 $\Omega_{\Lambda}$. Then it holds that 
 \begin{align} \label{eq:cgma}
  \Delta^{i\tau}_R = {\sf U}_{\beta \tau} \quad \ \ \text{with} \quad \ \ \beta = 2\pi/\kappa_*\, \quad \ \ (\tau \in \mathbb{R})
 \end{align}
\item[(4)] The previous relation \eqref{eq:cgma} can equivalently be expressed as stating that the state $\omega_\Lambda$ 
restricted to the $C^*$-subalgebra $\mathcal{W}_R = \mathcal{W}((-\infty,0) \times S_*)$ of $\mathcal{W}(\mathcal{D}_{S_*},\varsigma)$ is
a KMS-state for the action of the $\alpha_\tau$ at inverse temperature $\beta = 2\pi/\kappa_*$. Analogously, $\omega_\Lambda$
restricted to $\mathcal{W}_L = \mathcal{W}((0,\infty) \times S_*)$ is a KMS state for the action of the $\alpha_\tau$ at 
inverse temperature $\beta = - 2\pi/\kappa_*$. 
\end{itemize}

\subsection{Thermal interpretation of the 2-point function $\Lambda$} \label{se:thermality}

We will now point out that the thermal properties expressed in (3) and (4) at the end of the previous subsection can be directly 
read off from the Fourier spectrum of $\Lambda$ with respect to the Kodama time parameter, analogously as in \cite{MP}. 

To this end we recall the results presented in Section \ref{se:kodamaflow} and the action of the Kodama flow on the scaling limit state discussed in Section \ref{se:Kodama-Flow-Projected}.
In particular, points of $\mathcal{T}_*$
outside the Horizon $\mathcal{H}$ can be parametrized by $(u,\nn)$ where here the coordinate $u$ is related to $U$ by the following coordinate transformation
\begin{equation}\label{eq:change-U-u-coordinates}
U=-{\rm e}^{-\kappa_* u }, \quad U<0.
\end{equation}
In particular, we have seen in Proposition \ref{prop:scaled-kodama-flow} that the Kodama flow in the scaling limit, described by $ {\rm S}_{\tau}$,  acts as $u$-translation,
$u \mapsto u + \tau$. 
Thus, if $\varphi$ and $\varphi'$ are both contained in $C_0^\infty((-\infty,0) \times S_*,\mathbb{R})$, i.e.\ they are supported on $U < 0$, one obtains
\begin{equation}
\Lambda(\varphi,\varphi') = \lim_{\epsilon\to0^+} - \frac{r_*^2\kappa_*^2}{{4\pi}}
\int \frac{\varphi(u,\nn) \varphi'(u',\nn)}{\sinh\left((u-u')\frac{\kappa_*}{2}+i\epsilon\right)^2} \, du\,du'\, d\Omega^2(\nn)\,.
\end{equation}
A similar relation holds if $\varphi$ and $\varphi'$ are both supported on $U > 0$, on using the coordinate transformation $U = {\rm e}^{\kappa_* u}$. 
The Fourier transform along $u-u'$ of that distribution can be directly computed, see e.g.\ the Appendix of \cite{DMP}; it yields 
\begin{align} \label{Bose-dis}
 \Lambda(\varphi,\varphi') =  2 r_*^2
\int \frac{\overline{\hat{\varphi}(E,\nn)} \hat{\varphi}'(E,\nn)}{1 - {\rm e}^{-\beta E} } E\, dE\, d\Omega^2(\nn)\,, \quad \beta = 2\pi/\kappa_*\,,
\end{align} 
if $\varphi$ and $\varphi'$ are both supported either on $U < 0$ or $U >0$, where the Fourier transform with respect to $u$ has been denoted by a hat,
\begin{align}
 \hat{\varphi}(E,\nn) = \frac{1}{\sqrt{2 \pi}} \int {\rm e}^{-iEu} \varphi(u,\nn)\,du \,.
\end{align}
The appearance of the Bose thermal distribution factor $(1 - {\rm e}^{-\beta E})^{-1}$ for the Fourier ``energies'' in the integral expression \eqref{Bose-dis} 
manifestly shows the thermal Fourier spectrum of the 2-point function for an observer moving along the Kodama flow, where the inverse temperature is given by $\beta = 2\pi/\kappa_*$.

\subsection{Tunneling probability}

Again proceeding as in \cite{MP}, we now look at the Fourier transformed expression for the 2-point function $\Lambda$ in the case that $\varphi$ is supported on $U < 0$, i.e.\ outside of the 
outer trapping horizon, while $\varphi'$ is supported on its inside, on $U>0$. The result is (cf.\ \cite{MP}, Sec. 3.3 b)\,) 
\begin{align} \label{horizon-crossing}
\Lambda(\varphi,\varphi') =  {r_*^2}
\int \frac{\overline{\hat{\varphi}(E,\nn)} \hat{\varphi}'(E,\nn)}{\sinh(\frac{\beta}{2}E)} E\, dE\, d\Omega^2(\nn)\,.
\end{align}
This formula is the basis for estimating the tunneling probability or rather, transition probability between a one-particle state inside, and another one outside of the outer trapping horizon $\mathcal{H}$ in the scaling limit.
To this end, we choose some $E_0 > 0$, and define, for small $a > 0$, $\hat{\eta}_a(E) = 1$ if $|E - E_0| < a$, and $\hat{\eta}_a(E) = 0$ otherwise (i.e.\
$\hat{\eta}_a$ is the characteristic function of an interval of width $2a$ around $E_0$). We furthermore choose any non-zero, real, integrable, bounded function $b$ on $S_*$ and define 
\begin{align}
 \hat{h}_a(E,\nn) = \frac{1}{\sqrt{2 a}}\hat{\eta}_a(E)b(\nn)\,, \quad \ \ \hat{\varphi}_a(E,\nn) = \frac{\hat{h}(E,\nn)}{|| h_a ||_{(\Lambda)}}
\end{align}
where 
\begin{align}
 || h_a ||^2_{(\Lambda)} = \Lambda(h_a,h_a) =  2 r_*^2
\int \frac{|\hat{h}_a(E,\nn)|^2}{1 - {\rm e}^{-\beta E} } E\, dE\, d\Omega^2(\nn)\,, \quad \beta = 2\pi/\kappa_*\,,
\end{align}
defines the one-particle norm on $\mathcal{D}_{S_*}$ which is evidently finite for any $\hat{h}_a$. Therefore, $h_a$ (the inverse Fourier transform of $\hat{h}_a$) defines an 
element in $\overline{\mathcal{D}_{S_*}}^{(\Lambda)}$, the completion of $\mathcal{D}_{S_*}$ with respect to the norm $||\,.\,||_{(\Lambda)}$, supported on $U < 0$, the outside of $\mathcal{H}$.
Consequently $\varphi_a$, the inverse Fourier transform of $\hat{\varphi}_a$, is an element of $\overline{\mathcal{D}_{S_*}}^{(\Lambda)}$ which is supported on $U < 0$ and which is 
normalized, $|| \varphi_a ||_{(\Lambda)} = 1$. An $a$-parametrized family $\varphi'_{a}$ of elements in $\overline{\mathcal{D}_{S_*}}^{(\Lambda)}$ with 
$|| \varphi'_a ||_{(\Lambda)} = 1$, but supported on $U > 0$, is defined in complete analogy. We note that for each $a$, there is a sequence $\varphi_a^{(n)} \in \mathcal{D}_{S_*}$
$(n \in \mathbb{N})$ supported on $U < 0$ with $|| \varphi_a - \varphi_a^{(n)}||_{(\Lambda)} \underset{n \to \infty}{\longrightarrow} 0$. The same holds for primed counterparts of the 
functions involved, supported on $U > 0$. 

It also follows from the properties of the GNS representations of $\mathcal{W}$ for quasifree states that, defining ``one-particle vectors'' $\psi[\varphi_a^{(n)}] = -i\ln(\pi_\Lambda(W(\varphi_a^{(n)})))\Omega_\Lambda$
in $\mathscr{H}_\Lambda$, 
\begin{align}
 || \psi[\varphi_a^{(n)}] - \psi[\varphi_a^{(m)}] ||_{\mathscr{H}_\Lambda} = || \varphi_a^{(n)} - \varphi_a^{(m)} ||_{(\Lambda)}
\end{align}
holds. 
Therefore, the one-particle vectors $\psi[\varphi_a^{(n)}]$ form a Cauchy sequence, converging for any fixed $a$ to a unit vector, denoted by $\psi[\varphi_a]$, in $\mathscr{H}_\Lambda$.

We may now insert the expressions for $h_a$ and the analogously defined $h'_a$ into \eqref{Bose-dis} and \eqref{horizon-crossing}. Making use of the fact that 
$|\hat{h}_a|^2$, $|\hat{h}'_a|^2$ and $\overline{\hat{h}_a} \hat{h}'_a$ are delta-sequences with respect to $E$ peaked at $E_0$ as $a \to 0$, one finds in the limit 
$a \to 0$ for the transition probability
\begin{align}
\lim_{a \to 0}\, | \langle \psi[\varphi_a] , \psi[\varphi'_a] \rangle|^2 & = \lim_{a \to 0} \,\left|\Lambda(\varphi_a,\varphi'_a)\right|^2
= \frac{1}{4}\left(\frac{1 - {\rm e}^{-\beta E_0}}{\sinh(\beta E_0/2)}\right)^2 = {\rm e}^{-\beta E_0}.
\end{align}
For large enough values of $\beta E_0$, this approaches the Boltzmann thermal distribution since, if e.g.\ $\beta E_0 \ge \ln(2)$, then
\begin{align}
 \left| \frac{{\rm e}^{-\beta E_0}}{1 - {\rm e}^{-\beta E_0}} - {\rm e}^{-\beta E_0} \right| \le 2 {\rm e}^{-2 \beta E_0}
\end{align}
showing that in the limit of large $E_0$, the transition probability becomes exponentially suppressed as is characteristic of a thermal occupation of energy levels at inverse temperature 
$\beta = 2 \pi / \kappa_*$.
This observation is in agreement with  \cite{Zerbini1,Zerbini2,Zerbini3}.

\subsection{Coherent states of the scaling-limit-theory and their relative entropy}

Given the state $\omega_\Lambda$, its associated {\it coherent states} are of the form
\begin{align}
 \omega_{\varphi}(A) = \omega_\Lambda(W(\varphi)^* A W(\varphi))  \quad \ \ (A \in \mathcal{W}(\mathcal{D}_{S_*},\varsigma))\,.
\end{align}
This definition applies, in the first place, for all $\varphi \in \mathcal{D}_{S_*}$, but it is easy to see that it may be 
extended to all functions which lie in the completion $\overline{\mathcal{D}_{S_*}}^{(\Lambda)}$ of 
$\mathcal{D}_{S_*}$ with respect to the norm $|| \varphi ||_{(\Lambda)} = \Lambda(\varphi,\varphi)^{1/2}$. It is also 
easy to check that $\overline{\mathcal{D}_{S_*}}^{(\Lambda)}$ contains, e.g., $C_0^\infty(\mathbb{R},\mathbb{R}) \otimes 
 L^2_{\mathbb{R}}(S_*,d\Omega^2)$ (algebraic tensor product without completion). 
There is a particular feature shared by all coherent states: They are {\it completely uncorrelated with respect to 
the spatial (i.e.\ spherical) degrees of freedom}. This means, if there are finitely many subsets 
$G_j = I_j \times \Sigma_j$ ($j = 1,\ldots,n;\ n \ge 2)$ where the $I_j$ are real open intervals (admitting  the full real line) and 
the $\Sigma_j$ are open subsets of $S_* \simeq S^2$ which are pairwise disjoint, $\Sigma_j \cap \Sigma_k = \emptyset$ if 
$j \ne k$, then 
\begin{align} \label{eq:corr-free}
 \omega_\varphi(A_1 A_2 \cdots  A_n) = \omega_\varphi(A_1) \cdot \omega_\varphi(A_2) \cdots \omega_\varphi(A_n) 
\end{align}
holds for all $A_j \in \mathcal{W}(G_j)$. This relation generalizes to the case that $A_j \in \mathcal{N}(G_j)$, on extending
$\omega_\Lambda$ in the GNS representation to $\mathcal{B}(\mathscr{H}_\Lambda)$ as $\omega_\Lambda(B) = 
\langle \Omega_\Lambda,B\Omega_\Lambda \rangle$ $(B \in \mathcal{B}(\mathscr{H}_\Lambda))$. Note that $\omega_\Lambda$ itself 
is a coherent state (corresponding to $\varphi = 0$). 

For coherent states, the {\it relative entropy} can be easily calculated.
Without going into full details at his point, the relative entropy of a faithful, normal state on a von Neumann algebra with respect to another faithful, normal state was introduced 
by Araki \cite{Araki1} (see also \cite{Uhlmann}). It is a concept with an information theoretic background, see e.g.\ \cite{Donald,OhyaPetz} for 
further discussion. If $\omega_\varphi$ is any coherent state on $\mathcal{W}(\mathcal{D}_{S_*},\varsigma)$ as just described,
then in the GNS representation $(\mathscr{H}_\Lambda,\pi_\Lambda,\Omega_\Lambda)$ it is induced by the unit vector
$\Omega_\varphi = \pi_\Lambda(W(\varphi))\Omega_\Lambda$. If $\varphi$ is compactly supported in $(-\infty,0) \times S_*$ so that 
$\pi_\Lambda(W(\varphi))$ is contained in $\mathcal{N}_R$, then it is not difficult to see that $\Omega_\varphi$ is a standard vector 
for $\mathcal{N}_R$, meaning that $\Omega_\varphi$ is cyclic and separating for $\mathcal{N}_R$. In this case, the definition
of relative entropy in the sense of Araki applies for any pair of coherent states.
In particular, the relative entropy of $\omega_\varphi$ with respect to $\omega_\Lambda$ on $\mathcal{N}_R$ is given as
\cite{LongoECE,HollandsRECS}
\begin{align}
 S(\omega_\Lambda | \omega_\varphi) =  i \left. \frac{d}{dt}\right|_{t = 0}\langle \Omega_\varphi,\Delta_R^{it} \Omega_\varphi \rangle 
\end{align}
where $\Delta_R$ is, as above, the modular operator with respect to $\mathcal{N}_R$ and $\Omega_\Lambda$. 

To calculate $S(\omega_\Lambda|\omega_\varphi)$ in the case at hand (cf.\ again \cite{LongoECE,HollandsRECS} for similar calculations), we use \eqref{eq:cgma} to obtain
\begin{align}
 S(\omega_\Lambda|\omega_\varphi)= i \left. \frac{d}{dt}\right|_{t = 0}\langle \Omega_\varphi,\Delta_R^{it} \Omega_\varphi \rangle = i \left. \frac{d}{dt}\right|_{t = 0}\langle \Omega_\varphi, \Omega_{\varphi^t} \rangle 
\end{align}
where 
\begin{align}
 \varphi^t(U,\nn) = ({\sf S}_{2\pi t/\kappa_*}\varphi)(U,\nn) = \varphi({\rm e}^{2\pi t}U,\nn) \,.
\end{align}
Then we observe 
\begin{align}
 \langle \Omega_\varphi,\Omega_{\varphi^t}\rangle & = \omega_\Lambda(W(-\varphi)W(\varphi^t)) = {\rm e}^{\frac{i}{2}\varsigma(\varphi,\varphi^t)}\omega_\Lambda(W(\varphi^t - \varphi)) \nonumber  \\
 & = {\rm e}^{{\frac{i}{2}}\varsigma(\varphi,\varphi^t)}{\rm e}^{-\Lambda(\varphi^t - \varphi,\varphi^t - \varphi)/2}
\end{align}
Now we note that 
\begin{align}
 \left. \frac{d}{dt}\right|_{t = 0}\Lambda(\varphi^t - \varphi,\varphi^t - \varphi) 
  = 
 \left. \left(\Lambda(\mbox{$\frac{d}{dt}$}(\varphi^t - \varphi),\varphi^t - \varphi) + \Lambda(\varphi^t- \varphi,
 \mbox{$\frac{d}{dt}$} (\varphi^t - \varphi)) \right)\right|_{t = 0} = 0 
\end{align}
since $\varphi^t|_{t = 0} = \varphi$. 
Hence, we find
\begin{align} \label{eq:relent}
 S(\omega_\Lambda|\omega_\varphi) & = \left. { \frac{1}{2}}\frac{d}{dt}\right|_{t = 0} \varsigma(\varphi^t,\varphi) 
 = {-2 }\pi r_*^2 \int_{(-\infty,0)\times S^2} U (\partial_U \varphi)^2(U,\nn) \,dU\,d\Omega^2(\nn).
\end{align} 
In order to relate this entropy with the energy content of the coherent state measured by an observer moving along the Kodama flow, 
we rewrite the relative entropy formula with respect to the coordinate \eqref{eq:change-U-u-coordinates}. 
We then obtain 
\begin{equation}\label{eq:entropy-formula}
 S(\omega_\Lambda|\omega_\varphi)  
  = \beta {\mathcal{E}_\varphi}
\end{equation}
where 
\begin{equation}
 {\mathcal{E}_\varphi} = r_*^2 \int_{\mathbb{R}\times S^2}  (\partial_u \varphi)^2(u,\nn) \,du\,d\Omega^2(\nn)
\end{equation}
is the energy content of the coherent state $\omega_\varphi$ measured by the Kodama observer  
and $\beta=\frac{2\pi}{\kappa}$ is the inverse temperature of the KMS state $\omega_\Lambda$ (cf.\
Sec.\ 6.4 in \cite{KayWald}).

\subsection{Relative entropy is proportional to outer trapping horizon surface area}\label{se:rel-entropy-area}

The previous equality \eqref{eq:relent} establishes a proportionality between the relative entropy 
of coherent states of the scaling-limit-theory and the cross-section $S_*$ of the outer trapping horizon,
having the geometrical area $4\pi r_*^2$, with respect to which the scaling limit and the restriction to 
$\mathcal{T}_*$ of the quantized scalar field on the ambient spacetime are taken. This is justified if, for 
different such cross-sections, say $S_{*1}$ and $S_{*2}$, with respective radii $r_{*1}$ and $r_{*2}$, the associated
coherent states $\omega_\Lambda$ and $\omega_\varphi$ are identified. This is certainly very natural for the 
scaling limit state $\omega_\Lambda$, but for $\omega_\varphi$ that may, at first sight, not appear compelling. 
Let us therefore provide further motivation why the proportionality between the relative entropy of coherent states and
the surface area of the cross-section of the outer trapping horizon at which the scaling-limit-theory is considered 
arises naturally. The key point lies in the fact that the coherent states in the scaling-limit-theory are completely
correlation-free across spatial (i.e.\ spherical) separation as expressed in \eqref{eq:corr-free}, together with 
the additivity of the relative entropy with respect to correlation-free states.

To discuss this in more detail, fix a horizon cross-section $S_*$ with radius $r_*$ and consider the 
corresponding scaling limit Weyl-algebra $\mathcal{W}(\mathcal{D}_{S_*},\varsigma)$ with the scaling limit state
$\omega_\Lambda$, its GNS representation $(\mathscr{H}_\Lambda,\pi_\Lambda,\Omega_\Lambda)$ and the von Neumann algebras
$\mathcal{N}(G)$ for open subsets $G$ of $\mathbb{R}\times S_*$ as introduced in Sec.\ \ref{se:SLtheory}. Specifically,
for open subsets $\Sigma$ of $S_* \simeq S^2$, we define the von Neumann algebras
\begin{align}
 \mathcal{N}_R(\Sigma) = \mathcal{N}((-\infty,0) \times \Sigma).
\end{align}
We recall that $\mathcal{N}_R(\Sigma)$ is the von Neumann algebra contained in $\mathcal{B}(\mathscr{H}_\Lambda)$ generated 
by the $\pi_\Lambda(W(\varphi))$ where ${\rm supp}(\varphi) \subset (-\infty,0)$. Hence, on account of 
\eqref{eq:cgma}, the $\mathcal{N}_R(\Sigma)$ are invariant under the adjoint action of the modular group
$\Delta^{it}_R$ $(t \in \mathbb{R})$ with respect to $\mathcal{N}_R$ and $\Omega_\Lambda$. 

When we denote by $\omega_{\varphi,\Sigma}$ the state on
$\mathcal{N}_R(\Sigma)$ given by
\begin{align}
 A \mapsto \omega_{\varphi,\Sigma}(A) = \langle \Omega_\varphi,A \Omega_\varphi \rangle \quad 
 (A \in \mathcal{N}_R(\Sigma))\,,
\end{align}
i.e.\ the restriction of the coherent state $\omega_\varphi$ defined previously to $\mathcal{N}_R(\Sigma)$,
and if likewise the restriction of $\omega_\Lambda$ to $\mathcal{N}_R(\Sigma)$ is denoted by $\omega_{\Lambda,\Sigma}$,
then we find for the relative entropy in the same way as before,
\begin{align}
 S(\omega_{\Lambda,\Sigma}|\omega_{\varphi,\Sigma})  = -{2}\pi r_*^2 \int_{(-\infty,0)\times S^2} U (\partial_U \varphi)^2(U,\nn) \,dU\,d\Omega^2(\nn) = S(\omega_\Lambda|\omega_\varphi)\,.
\end{align}
Then, if $\Sigma_1$ and $\Sigma_2$ are any two {\it disjoint} open subsets of $S^2$, and if 
${\rm supp}(\varphi_j) \subset (-\infty,0) \times \Sigma_j$ ($j = 1,2$), and setting $\varphi_{12} = 
\varphi_1 + \varphi_2$, one finds 
\begin{align}
 S&(\omega_{\Lambda,\Sigma_1 \cup \Sigma_2}|\omega_{\varphi_{12},\Sigma_1 \cup \Sigma_2})
 = S(\omega_\Lambda|\omega_{\varphi_{12}}) \nonumber \\
 & = -{2}\pi r_*^2 \int_{(-\infty,0)\times S^2} U (\partial_U \varphi_{12})^2(U,\nn) \,dU\,d\Omega^2(\nn) 
 \nonumber \\
 & = -{2}\pi r_*^2 \int_{(-\infty,0)\times S^2} U \left[(\partial_U \varphi_1)^2(U,\nn)
 +  (\partial_U \varphi_2)^2(U,\nn) \right] \,dU\,d\Omega^2(\nn) 
 \nonumber \\
 & = S(\omega_\Lambda|\omega_{\varphi_1}) + S(\omega_\Lambda|\omega_{\varphi_2}) \nonumber \\
 & = S(\omega_{\Lambda,\Sigma_1}|\omega_{\varphi_1,\Sigma_1}) + S(\omega_{\Lambda,\Sigma_2}|\omega_{\varphi_2,\Sigma_2})
\end{align}
where we passed from the 3rd equality to the 4th since $\varphi_1$ and $\varphi_2$ are assumed to have 
disjoint $\nn$-supports. This shows that the relative entropy of coherent states in any scaling limit is {\it additive 
with respect to angular separation}; actually, a corresponding additivity of the relative entropy across 
angular separation holds upon replacing the two open, disjoint subsets $\Sigma_1$ and $\Sigma_2$ of $S^2$ by 
finitely many $\Sigma_1,\ldots,\Sigma_N$, and similarly $\varphi_1$ and $\varphi_2$ by finitely many $\varphi_1,\ldots,\varphi_N$ with ${\rm supp}(\varphi_j) \subset (-\infty) \times \Sigma_j$. 

In fact, this can be seen to be, more generally, a consequence of the fact that the coherent states in the scaling 
limit are correlation-free across angular separation, and the additivity of the relative entropy of correlation-free states.
One can show that there is a joint unitary equivalence $\omega_{\varphi_{12},\Sigma_1 \cup \Sigma_2} \simeq \omega_{\varphi_1,\Sigma_1} \otimes \omega_{\varphi_2,\Sigma_2}$ and 
$\omega_{\Lambda,\Sigma_1 \cup \Sigma_2} \simeq \omega_{\Lambda,\Sigma_1} \otimes \omega_{\Lambda,\Sigma_2}$,
where the {\it correlation-free product state} $\omega_{\varphi_1,\Sigma_1}\otimes \omega_{\varphi_2,\Sigma_2}$ is 
the state defined on $\mathcal{N}_R(\Sigma_1) \otimes \mathcal{N}_R(\Sigma_2)$  by linear extension of 
\begin{align}
 A_1 \otimes A_2 \mapsto \omega_{\varphi_1,\Sigma_1}\otimes \omega_{\varphi_2,\Sigma_2}(A_1 \otimes A_2) = 
 \omega_{\varphi_1,\Sigma_1}(A_1) \cdot \omega_{\varphi_2,\Sigma_2}(A_2) \,.
\end{align}
For (faithful, normal) correlation-free product states, the equation 
\begin{align}
 S(\omega_{\Lambda,\Sigma_1} \otimes \omega_{\Lambda,\Sigma_2} | \omega_{\varphi_1,\Sigma_1} \otimes \omega_{\varphi_2,\Sigma_2} ) = S(\omega_{\Lambda,\Sigma_1} | \omega_{\varphi_1,\Sigma_1}) + 
 S(\omega_{\Lambda,\Sigma_2} | \omega_{\varphi_1,\Sigma_2})
\end{align}
holds (cf.\ \cite{OhyaPetz}, eq.\ (5.22)), whereupon one may conclude that 
\begin{align}
 S(\omega_{\Lambda,\Sigma_1 \cup \Sigma_2}|\omega_{\varphi_{12},\Sigma_1 \cup \Sigma_2}) = 
 S(\omega_{\Lambda,\Sigma_1} | \omega_{\varphi_1,\Sigma_1}) + 
 S(\omega_{\Lambda,\Sigma_2} | \omega_{\varphi_1,\Sigma_2})
\end{align}
obtains.

Therefore, the scaling of the relative entropy of coherent states proportional to the geometric area of the horizon cross-section arises naturally.
This is seen particularly cleary when considering coherent states corresponding to elements $\varphi \in
\overline{\mathcal{D}_{S_*}}^{(\Lambda)}$ which are of the form $\varphi = h \odot \chi_{\Sigma}$ where 
\begin{align}
 (h \odot \chi_\Sigma)(U,\nn) = h(U) \cdot \chi_\Sigma(\nn)  \quad (U \in (-\infty,0)\,, \ \nn \in S^2)
\end{align}
with $h \in C_0^\infty((-\infty,0),\mathbb{R})$ and $\chi_{\Sigma}$ the characteristic function of an 
open, or more generally, measurable subset $\Sigma$ of $S^2$. In this case,
\begin{align}
S(\omega_\Lambda| \omega_{h \odot \chi_\Sigma}) & = -{2} \pi  \int_{-\infty}^0 U (\partial_U h)^2(U) \, dU \cdot 
 r_*^2\int_{\Sigma \subset S^2} d\Omega^2(\nn) \nonumber \\
 & = - {2} \pi \int_{-\infty}^0 U (\partial_U h)^2(U) \, dU \cdot \mathcal{A}(\Sigma_{r_*} \subset S_*) 
\end{align}
where $\mathcal{A}(\Sigma_{r_*} \subset S_*)$ is the geometrical surface of $\Sigma$ viewed as subset of 
the horizon cross-section $S_*$ which is a copy of the 2-dimensional sphere with radius $r_*$, i.e.\ the 
surface of $\Sigma$ as subset of $S^2$, scaled by $r_*^2$. 

In the light of these observations, it is entirely natural to identify, if $\Sigma = S^2$, the 
coherent states $\omega_{h \odot 1}$  for different horizon cross-sections $S_*$ with different radii $r_*$, 
which renders a proportionality of the relative entropies with the horizon cross-section area $\mathcal{A}(S_*)$,
\begin{align}
 S(\omega_\Lambda | \omega_{h \odot 1}) = -{2} \pi \int_{-\infty}^0 U (\partial_U h)^2(U) \, dU \cdot \mathcal{A}(S_*) 
\end{align}
for the coherent states of the said type, when considering the scaling-limit-theory taken at $S_*$, arising 
from any Hadamard state of the quantum field theory on the underlying spherically symmetric spacetime with 
an outer trapping horizon. 
\\[6pt]
{\bf Remark} \ \ Without the factor 2 in the definition of the scaling transformations ${\sf u}_\lambda$, one would obtain 
that the relative entropy $S(\omega_\Lambda | \omega_{h \odot 1})$ equals one quarter of the horizon cross-sectional area times 
$-{2} \pi \int_{-\infty}^0 U (\partial_U h)^2(U) \, dU$, where the latter is the relative entropy of the coherent state induced 
by $h$ of the free chiral conformal quantum field theory defined on the real line with the vacuum two-point function 
\begin{align}
 \Lambda_{(1)}(h,h') = 
\lim_{\varepsilon\to 0+} -\frac{1}{{\pi}} \int  \frac{ h(U) h'(U')}{(U-U'+i\varepsilon)^2}
dU\,dU' \quad \ \ (h,h' \in C_0^\infty(\mathbb{R},\mathbb{R}))
  \end{align}
This is in close analogy to the classical derivation where black hole entropy is equated to one quarter of the cross-sectional 
horizon area. 
Yet, it should be borne in mind that it refers not to the entropy of the outer trapping horizon itself but to 
quantities of a quantum field theory arising in the scaling limit towards a spherical cross-section of the outer trapping horizon.
Therefore, the value of the relative entropy depends on the states chosen and also on the field content of the initially considered 
quantum field theory. Nevertheless, regardless of such choices, there is a characteristic scaling of that relative entropy proportional to (one quarter of) the cross-sectional 
area of the outer trapping horizon with respect to which the scaling limit is considered.

\section{Conclusion} 
In this paper we have 
investigated the scaling limits of Hadamard 2-point functions on the lightlike submanifold $\mathcal{T}_*$ of a
spherically symmetric outer trapping horizon generated by lightlike geodesics traversing the outer trapping horizon.
The scaling limit 2-point function $\Lambda$ was found to have a universal form, independent of which Hadamard 2-point function
of the quantum field theory on the underlying spherically symmetric spacetime is initially chosen. The projected Kodama flow 
acts in the scaling limit like a dilation, and the scaling limit 2-point function $\Lambda$ shows a thermal spectrum with respect
to the projected Kodama flow at inverse temperature $\beta = 2\pi/\kappa_*$ where $\kappa_*$ is the surface gravity of the 
horizon cross-section $S_*$ where the lightlike generators of $\mathcal{T}_*$ traverse the outer trapping horizon.
Consequently, one can derive a tunneling probability in the scaling limit for Fourier modes peaked at 
Fourier energy $E_0$ with respect to the Kodama time behaving like ${\rm e}^{-\beta E_0}$ for large $E_0$, analogous to a thermal distribution of 
energy modes.
These results 
are in agreement with earlier, related results for stationary black horizons or bifurcate Killing horizons, in particular
\cite{MP} (see also \cite{KayWald,Zerbini1,Zerbini3}), and also with
the first law of non-stationary black hole dynamics discussed by Hayward \cite{Hayward},
$\mathcal{M}' = \frac{\kappa }{8\pi} \mathcal{A}' + w \mathcal{V}'$ mentioned in the Introduction. 

Furthermore, the scaling limit 2-point function $\Lambda$ defines a quantum field theory on each $\mathcal{T}_*$, the scaling-limit-theory, determined 
by the horizon cross-section $S_*$. The thermal Fourier spectrum with respect to the Kodama time in the scaling limit is equivalent to 
the KMS property of the scaling limit state $\omega_\Lambda$ induced by $\Lambda$ when restricted to observables localized on the part of $\mathcal{T}_*$ lying either
inside or outside of the outer trapping horizon. Furthermore, the state $\omega_\Lambda$ as well as all the coherent states $\omega_\varphi$ in the scaling-limit-theory 
are correlation-free product states with respect to separation in the angular coordinate $\nn$ of $S_*$, and we have seen that this leads naturally to a proportionality 
of the relative entropy $S(\omega_\Lambda | \omega_\varphi)$ with $4\pi r_*^2$, the area of the cross-section $S_*$ defining the scaling-limit-theory. 
Again, this is in keeping with the classical theory of black hole thermodynamics \cite{BardeenCarterHawking,Hayward,Bekenstein}. We emphasize that this 
is a consequence of our scaling limit analysis and seems to be the first such result in the setting of quantum field theory in curved spacetime (apart from
the related arguments of \cite{HollIshi}). 

We should remark that our scaling limit consideration is akin to an adiabatic limit in the sense that effectively, in the scaling limit all processes or dynamical changes 
at finite time-scales are being scaled away. In this sense, our concepts of inverse temperature and of relative entropy in the scaling-limit-theory are 
not dynamical, and that is a considerable limitation of our approach. The entropy concept, in our the scaling limit, bears some similarity to that in Bekenstein's
early article \cite{Bekenstein} on the subject: When an object (e.g.\ a table, a chair or a tankard) traverses the horizon, then the
information about the object is lost outside of the horizon. In \cite{Bekenstein}, the example of beams of light entering a black hole horizon is used. Our scaling-limit-theory can be seen
as a bunch of elementary theories for such beams of light, namely, a free chiral conformal field theory, one for each point on $S_*$. As the area of $S_*$ is 
increased, for example, it accomodates for more such ingoing light beams as measured by the area, and correspondingly a larger amount of information along ``elementary
light beams'' passing the horizon through $S_*$ can be 
absorbed, which corresponds to the scaling of entropy -- a measure for the lost amount of information -- proportional to the area of $S_*$. 
See also the article \cite{CLR} which can be regarded as a quantum version of Bekenstein's attribution of 
entropy to beams of light.

We think that similar results can also be obtained for other types of horizons, like cosmological horizons \cite{dangelo} or isolated horizons. 
A greater challenge is to attempt to obtain a more dynamical concept of temperature and entropy for dynamical black hole horizons in the setting of 
quantum field theory in curved spacetimes and semiclassical gravity, in the spirit of the approach of \cite{HollIshi} which takes dynamical metric perturbations
around a static Schwarzschild black hole horizon into account
(see, e.g., the recent work \cite{Dan21}). 
It would also be of interest to see if the notions of temperature and entropy in the context 
of our semiclassical approach to the temperature and entropy of black hole horizons can be linked to more ``holographic'' entropy concepts \cite{HuebRanTak}.

\subsection*{Acknowledgments}

F.K.\ thanks the IMPRS at the Max-Planck-Institute for Mathematics in the Sciences, Leipzig, for financial support.
N.P.\ thanks the ITP of the University of Leipzig for the kind hospitality during the preparation of this work and the DAAD for supporting this visit with the program ``Research Stays for Academics 2017". We would also like to thank Valter Moretti for discussion 
surrounding the definition of the Hadamard form. 

\section{Appendix} \label{Appendix}
In this Appendix, we present the proof of Thm.\ \ref{Thm:SL}. The proof will be facilitated by the following auxiliary result.
\begin{lemma} \label{lm:delta-estimate} 
Let $(\varrho_{\alpha})_{0 < \alpha_k \le 1}$  $(k = 1,\ldots, n)$ be a family of
measurable  functions $\varrho_{\alpha} : \mathbb{R}^m \to \mathbb{C}$, indexed by 
$\alpha = (\alpha_1,\ldots,\alpha_n) \in \mathbb{R}^n$, so that for any compact 
subset $C$ of $\mathbb{R}^m$ there is some $0< a_0 \le 1$ with
\begin{align}
\sup_{0 < \alpha_k \le a_0}\, \sup_{y \in C}\,| \varrho_{\alpha}(y) | \le \frac{1}{2} \quad 
\end{align}
Furthermore, let $(j_{\alpha})_{0 < |\boldsymbol{\lambda}| \le 1}$ be a family of continuous 
functions $j_{\alpha} : \mathbb{R}^m \to \mathbb{C}$, with the properties:
\begin{itemize}
 \item[(i)] ${\rm supp}(j_{\alpha}) \subset J^m$ with some
 fixed compact real interval $J$;
 \item[(ii)] $| j_{\alpha}(y)| \le b$ for all $\alpha$ and all 
 $y \in J^m$ with a fixed finite constant $b > 0$.
\end{itemize}
Then the following statements hold.
\\[6pt]
{\bf (A)} \quad
There are some some $0 < a_0 \le 1$ and a finite positive constant $\kappa$ so that, if $0 < \delta \le \frac{1}{2}$, it holds that
(writing $y = (y_1,\underline{y})$ and  $d^my = dy_1 \,d^{m-1}\underline{y}$)
\begin{align}
\sup_{0 < \alpha_k \le a_0}\, \int_{\mathbb{R}^{m-1}} \int_0^\delta \left| \ln|\varrho_{\alpha}(y) + y_1^2 |\right|\,|j_{\alpha}(y)
  | \,dy_1 d^{m-1} \underline{y} \ \le \ \kappa \delta^{1/3} \,.
\end{align}
{\bf (B)} \quad 
There are some $0 <a_0 \le 1$ and a finite constant $\kappa$ so that, if $\delta \le \frac{1}{2}$, it holds that 
(writing $y = (y_0,y_1,\underline{y})$ and $d^m y = dy_0 d y_1 d^{m-2}\underline{y})$)
\begin{align}
\sup_{0 < \alpha_k \le a_0}\, \int_{\mathbb{R}^{m-2}} \int_{ | y_0^2 - y_1^2| < \delta^2} \left| \ln|\varrho_{\alpha}(y) + y_0^2 - y_1^2 |\right|\,|j_{\alpha}(y)
  | \,dy_0\,dy_1 d^{m-2} \underline{y} \ \le \ \kappa \delta^{1/3} \,.
\end{align}
\end{lemma}
\noindent
{\it Proof of Lemma \ref{lm:delta-estimate} }\\[6pt]
Part {\bf (A)} \\[2pt]
Making use of the integration coordinate substitution $z = y_1^2$, thus $dz = 2 y_1 d y_1$, 
\begin{align}
\int_{\mathbb{R}^{m-1}} & \int_0^\delta \left| \ln|\varrho_{\alpha}(y) + y_1^2|\,j_{\alpha}(y)
  \right|\,d^my = 
 \int_{\mathbb{R}^{m-1}} \int_0^{\delta^2} \left| \ln|\varrho_{\alpha}(\sqrt{z},\underline{y}) + z |
 \frac{j_{\alpha}(\sqrt{z},\underline{y})}{2 \sqrt{z}}
  \right|\,dz\,d^{m-1}\underline{y}  \,.
\end{align}
H\"older's integral inequality with $p = 3$, $q = 3/2$ (so that $1/p + 1/q = 1$) with respect to the $z$-integration yields 
\begin{align}
 \int_{\mathbb{R}^{m-1}} & \int_0^{\delta^2} \left| {\ln|\varrho_{\alpha}(\sqrt{z},\underline{y}) + z |}
\frac{ j_{\alpha}(\sqrt{z},\underline{y})}{2 \sqrt{z}}
  \right|\,dz\,d^{m -1}\underline{y} \\
 & \le \frac{1}{2}\int_{\mathbb{R}^{m-1}} \left[ \int_0^{\delta^2} |\,\ln|\varrho_{\alpha}(\sqrt{z},\underline{y}) + z|\,|^3\,d z \right]^{1/3}
  \left[ \int_0^{\delta^2}  z^{-3/4} |j_{\alpha}(\sqrt{z},\underline{y})|^{3/2} \,dz\right]^{2/3} \,d^{m-1}\underline{y} \, \nonumber
\end{align}
Choosing $0 < a_0 \le 1$ so that 
\begin{align} \label{eq:a-null}
\sup_{0 < \alpha_k \le a_0}\, \sup_{y \in J^m}\,| \varrho_{\alpha}(y) | \le \frac{1}{2} \,,
\end{align}
then for $0 < \alpha_k \le a_0$, the last integral can be estimated by
\begin{align}
 \sup_{0 < |\varrho| \le 1/2} \frac{1}{2}\left[\int_0^{\delta^2}  |\, \ln| \varrho + z| \, |^3 \,dz \right]^{1/3} 
 \left[ \int_0^{\delta^2}  z^{-3/4} b^{3/2}\,dz\right]^{2/3} \, |J|^{m -1}
\end{align}
where $|J|$ is the length of the interval $J$.

We observe that since $|\varrho| \le 1/2$ and $0 < z \le \delta^2$ with $\delta \le 1/2$, we obtain
$|{\rm Re}(\varrho) + z|^2 + |{\rm Im}(\varrho)|^2 = | \varrho + z|^2 < 1$. Consequently, under the integral,
$|\,\ln| \varrho + z | \,| \le | \, \ln | {\rm Re}(\varrho) + z |\,|$. This results 
in 
\begin{align}
 \sup_{0 < |\varrho| \le 1/2} \left[\int_0^{\delta^2}  |\, \ln| \varrho + z| \, |^3 \,dz \right]^{1/3} 
 & \le \sup_{0 < |\varrho| \le 1/2} \left[\int_0^{\delta^2}  |\, \ln| {\rm Re}(\varrho) + z| \, |^3 \,dz \right]^{1/3} 
 \nonumber \\
 & = \sup_{0 < |r| \le 1/2}  \left[\int_{-r}^{\delta^2 - r }  |\, \ln| z| \, |^3 \,dz \right]^{1/3} 
  \le K
\end{align}
with a finite, positive real constant $K$. On the other hand, we obtain 
\begin{align}
 \left[ \int_0^{\delta^2}  z^{-3/4} \,dz\right]^{2/3} = 4^{2/3} \delta^{1/3}\,.
\end{align}
Putting all the previous steps together, we find
\begin{align}
 \sup_{0 < \alpha_k \le a_0}\, \int_{\mathbb{R}^{m-1}} \int_0^\delta \left| \ln|\varrho_{\alpha}(y) + y_1^2|\right|\,|j_{\alpha}(y)
  |\, dy_1 d^{m-1}\underline{y} \le  \frac{4^{2/3}}{2}b|J|^{m-1} K \delta^{1/3}\,.
\end{align}
This proves the statement of Part {\bf (A)}, with $\kappa =  \frac{4^{2/3}}{2}b |J|^{m-1} K$. 
\\[6pt]
Part {\bf (B)} \\[2pt]
First we note that the set $| y_0^2 - y_1^2| < \delta^2$ in the $y_0$-$y_1$-plane can be split into the four parts
\begin{align}
 H_1(\delta) & = \{ |y_1| \le y_0 < \sqrt{y_1^2 + \delta^2}\} \,, \quad \ \ H_2(\delta) = \{ - \sqrt{y_1^2 + \delta^2} < y_0 \le - |y_1|\} \\
 H_3(\delta) & = \{ |y_0| \le y_1 < \sqrt{y_0^2 + \delta^2}\} \,, \quad \ \ H_4(\delta) = \{ - \sqrt{y_0^2 + \delta^2} < y_1 \le - |y_0|\}
\end{align}
The sets overlap only at their boundaries, $y_0 \pm y_1 =0$. Therefore,
\begin{align}
\int_{\mathbb{R}^{m-2}} & \int_{ | y_0^2 - y_1^2| < \delta^2} \left| \ln|\varrho_{\alpha}(y) + y_0^2 - y_1^2 |\right|\,|j_{\alpha}(y)
  | \,dy_0\,dy_1 \,d^{m-2} \underline{y} \nonumber \\
 &  = \sum_{\ell = 1}^4 \int_{\mathbb{R}^{m-2}} \int_{H_\ell(\delta)} \left| \ln|\varrho_{\alpha}(y) + y_0^2 - y_1^2 |\right|\,|j_{\alpha}(y)
  | \,dy_0\,dy_1\, d^{m-2} \underline{y}
\end{align}
The integrals involving the $H_\ell(\delta)$ all have a very similar structure and thus it suffices to show that, e.g.,
\begin{align}
\sup_{0< \alpha_k \le a_0} \int_{\mathbb{R}^{m-2}} \int_{H_1(\delta)} \left| \ln|\varrho_{\alpha}(y) + y_0^2 - y_1^2 |\right|\,|j_{\alpha}(y)
  | \,dy_0\,dy_1 \, d^{m-2} \underline{y} \ \,  \le \ \, \kappa_1 \delta^{1/3}
\end{align}
since similar estimates for the other $H_\ell(\delta)$ can be deduced by analogous arguments.
Carrying out a substitution $z = y_0^2$ followed by a H\"older-type integral inequality similarly as in the proof of Part {\bf (A)},
we find, on making $a_0$ small enough so that \eqref{eq:a-null} holds, for all $0 < \alpha_k \le a_0$,
\begin{align}
 & \int_{\mathbb{R}^{m-2}} \int_{H_1(\delta)} \left| \ln|\varrho_{\alpha}(y) + y_0^2 - y_1^2 |\right|\,|j_{\alpha}(y)
  | \,dy_0\,dy_1 \,d^{m-2} \underline{y}  \nonumber \\
& = \int_{\mathbb{R}^{m-2}} \int_J \int_{|y_1|}^{\sqrt{y_1^2 + \delta^2}} \left| \ln|\varrho_{\alpha}(y) + y_0^2 - y_1^2 |\right|\,|j_{\alpha}(y)
  | \,dy_0\,dy_1 \,d^{m-2} \underline{y} \nonumber \\
& \le \frac{1}{2}\int_{\mathbb{R}^{m-2}} \int_J \left[\int_{y_1^2}^{y_1^2 + \delta^2} \left| \ln|\varrho_{\alpha}(\sqrt{z},y_1,\underline{y}) + z - y_1^2 |\right|^3 \,dz \right]^{1/3} \times \\
& \quad \quad \quad \quad \quad  \times \left[ \int_{y_1^2}^{y_1^2 + \delta^2} \frac{|j_\alpha(\sqrt{z},y_1,\underline{y})|^{3/2}}{z^{3/4}} \,dz \right]^{2/3}
 dy_1\,d^{m -2}\underline{y} \nonumber
 \\
& \le \frac{1}{2} \sup_{|\varrho| < 1/2}\left[\int_{0}^{\delta^2} \left| \ln| \varrho + z  |\right|^3 \,dz \right]^{1/3} b |J|^{m-1} 
\sup_{y_1 \in J} \left[ \int_{y_1^2}^{y_1^2 + \delta^2} z^{-3/4} \,dz \right]^{2/3} \label{eq:delta-ln-2}
\end{align}
where an obvious substitution of $z$ by $z -y_1^2$ has been carried out in the integral involving the logarithm. It is easy to check 
that 
\begin{align}
 \sup_{y_1 \in J}\left[ \int_{y_1^2}^{y_1^2 + \delta^2} z^{-3/4} \,dz \right]^{2/3} \le  \left[ \int_{0}^{\delta^2} z^{-3/4} \,dz \right]^{2/3}
\end{align}
and therefore we see that the integral expression in \eqref{eq:delta-ln-2}
can be estimated by 
\begin{align}
 \frac{4^{2/3}}{2}b|J|^{m-1} K \delta^{1/3} 
\end{align}
just as in the proof of Part {\bf (A)}. This concludes the proof of Part {\bf (B)} \hfill $\Box$
\\ \\
{\it Proof of Theorem \ref{Thm:SL}.}
It will be convenient to introduce the following abbreviations, referring to adapted coordinates $(U,V,\nn)$ near the chosen $S_*$:
\begin{align}
 x = & (U,V,\nn)\,, \quad x' = (U',V',\nn')\,, \quad x_\lambda = (\lambda U,V,\nn)\,, \quad x'_\lambda = (\lambda U',V',\nn')\,, \\
  & dX = dX(x) = dU\,dV \,d\Omega^2(\nn) = dU\,dV\,\sin (\vartheta)d\vartheta\,d\varphi 
\end{align}
using $\nn = (\vartheta,\varphi)$ in spherical angular coordinates as before; $dX'$ is defined analogously.
Another abbreviations that we will use are
\begin{align}
 \tilde{F}_\lambda(U,V,\nn) = \eta^{-3}(U,V)F_\lambda(U,V,\nn)\, , \quad \ \ F_\lambda(U,V,\nn) = ({\sf u}_\lambda {2}\partial_U f)(U,V,\nn) 
\end{align}
and analogously for symbols endowed with primes.

Recalling \eqref{eq:match}, we have 
\begin{align}
 w^{(2)}(F_\lambda,F'_\lambda) & = \tilde{w}{}^{(2)}(\tilde{F}_\lambda,\tilde{F}'_\lambda)  \nonumber \\
                               & = \lim_{\varepsilon \to 0+}\, \int \tilde{w}_\varepsilon(x,x') \tilde{F}_\lambda(x)
                               \tilde{F}'_\lambda(x') \,d{\rm vol}_{\tilde{g}}(x) \,d{\rm vol}_{\tilde{g}}(x') \nonumber \\
                               & = \lim_{\varepsilon \to 0+} \int \tilde{w}_\varepsilon(x,x') F_\lambda(x) F'_\lambda(x')
                                P(U,V,U',V')\,dX\,dX'
\end{align}
having made use of $d{\rm vol}_{\tilde{g}}(x) = \eta(U,V)^2A(U,V)r_*^2\,dX$ in the adapted coordinates for $S_*$ which, as we recall, is 
a copy of a 2-sphere with radius $r_*$. 
We introduce on a smooth partition of 
unity on $S^2 \times S^2$, consisting of two functions $\chi$ and $\chi^\perp$, as follows:  
Choose some $0 < \delta^2 < \pi^2/64$ and choose a non-negative $C^\infty$ function $\chi$, bounded by 1, on $S^2 \times S^2$ so that 
$\chi(\nn,\nn') = 1$ if ${\sf s}(\nn,\nn')/r_*^2 \le \delta^2/2$, and $\chi(\nn,\nn') = 0$ if ${\sf s}(\nn,\nn')/r_*^2 \ge \delta^2$. We then write 
$\chi^{\perp}(\nn,\nn') = 1 - \chi(\nn,\nn')$. Note that $\chi = \chi_\delta$ and $\chi^\perp = \chi_\delta^\perp$ depend on the choice of $\delta$.
With this notation, we can write 
\begin{align}
 \tilde{w}_{\varepsilon}(x,x') = \tilde{w}_{\varepsilon}(x,x')\chi(\nn,\nn') + \tilde{w}_\varepsilon(x,x')\chi^\perp(\nn,\nn') \,.
\end{align}
In a further step, we observe that, on a change of the $U$ and $U'$ integration coordinates,
\begin{align} \label{eq:chiperp}
 \int & \tilde{w}_\varepsilon(x,x')\chi^\perp(x,x') F_\lambda(x) F'_\lambda(x')
                                P(U,V,U',V')r_*^2 \,dX\,dX'\\
 & = \int \left[\tilde{\psi}(x_\lambda,x'_\lambda) \left( \frac{1}{8\pi^2}\frac{\tilde{\Delta}{}^{1/2}(x_\lambda,x'_\lambda)}{\tilde{\sigma}_\varepsilon(x_\lambda,x'_\lambda)} + 
\ln(\tilde{\sigma}_\varepsilon(x_\lambda,x'_\lambda))\tilde{Y}(x_\lambda,x'_\lambda) \right) + \tilde{Z}(x_\lambda,x'_\lambda)\right] \times  \nonumber \\
& \quad \quad \quad \quad \quad \times
P(\lambda U,V,\lambda U',V')\chi^\perp(\nn,\nn')F(x)F(x')\,dX\,dX'\,. \nonumber 
\end{align}
In view of the particular form of the half of the squared geodesic distance \eqref{eq:pitagora}, we have 
\begin{align}
 \tilde{\sigma}_\varepsilon(x_\lambda,x_\lambda') = \tilde{\sigma}^{(\mathcal{L})}(\lambda U,V,\lambda U',V') + {\sf s}(\nn,\nn') + 2i\varepsilon t(\lambda U,V,\lambda U',V') +  \varepsilon^2
\end{align}
Since in the integral on the right hand side of \eqref{eq:chiperp}, ${\sf s}(\nn,\nn') \ge \delta > 0$ owing to the presence of $\chi^\perp$, the integrand functions remain 
uniformly bounded in the limits as $\varepsilon \to 0$ and $\lambda \to 0$, and they converge almost everywhere to an integrable function. Therefore, one obtains 
\begin{align}
\lim_{\lambda \to 0+} & \lim_{\varepsilon \to 0+}  \int  \tilde{w}_\varepsilon(x,x')\chi^\perp(x,x') F_\lambda(x) F'_\lambda(x')
                                P(U,V,U',V') \,dX\,dX' \nonumber \\
                              &  = \int h(V,V',\nn,\nn') \partial_U f(U,V,\nn) \partial_{U'}f'(U',V',\nn')\,dX \,dX' = 0
\end{align}
for some bounded $L^1$ function $h$; the integral on the right hand side vanishes since, after the limit $\lambda \to 0$, $h$ is independent of $U$ and $U'$, and 
$f$ and $f'$ have compact support (in particular, compact support with respect to $U$, respectively $U'$). Note that this holds no matter how small $\delta > 0$ has been
chosen.

Next we notice 
that ${\sf s}(\nn,\nn')$ is invariant under rotations $R \in SO(3)$, ${\sf s}(R\nn,R\nn') = {\sf s}(\nn,\nn')$;
similarly,
the surface-integration form $d\Omega^2$ is invariant unter the rotations, $d\Omega^2(R\nn) = d\Omega^2(\nn)$. Therefore, given $\nn'$,
we can regard it as obtained from a standard ``north pole point'' $\overset{\circ}{\nn}{}'$ by a suitable rotation $R_{\nn'} \in SO(3)$
so that $\nn' = R_{\nn'}\overset{\circ}{\nn}{}'$, hence  ${\sf s}(\nn,\nn') = {\sf s}(\nn,R_{\nn'}\overset{\circ}{\nn}{}') = {\sf s}(R_{\nn}^{-1}\nn,\overset{\circ}{\nn}{}')$.
The relation between $\nn'$ and $R_{\nn'}$ is bijective and smooth 
as long as $\nn'$ is bounded away by a finite distance from the antipode point $- \overset{\circ}{\nn}{}'$ (on identifying 
$\overset{\circ}{\nn}{}' = (0,0,1) \in \mathbb{R}^3$). In the following integrals we will consider this is always the case owing to 
the presence of the function $\chi(\nn,\nn')$. 
Introducing the abbreviations
\begin{align}
  \tilde{\sigma}_{[\lambda]}^{(\mathcal{L})} = \tilde{\sigma}^{(\mathcal{L})}(\lambda U,V,\lambda U',V')\,, \quad t_{[\lambda]} = t(\lambda U,V, \lambda U',V')\,,
\end{align}
we thus have, for any $\lambda$-parametrized family $q_\lambda(x,x')$ of bounded, compactly supported $C^1$ functions, writing
$R_{\nn'}x= (U,V,R_{\nn'}\nn)$ for $x = (U,V,\nn)$
\begin{align}
\int \frac{q_\lambda(x,x')}{\tilde{\sigma}_\varepsilon(x_\lambda,x'_\lambda)}\chi(\nn,\nn') \,dX\,dX' \label{eq:qlambda} 
 & = \int \frac{q_\lambda(x,x')}{\tilde{\sigma}_{[\lambda]}^{(\mathcal{L})} + {\sf s}(R_{\nn'}^{-1}\nn,\overset{\circ}{\nn}{}') + 2i\varepsilon t_{[\lambda]} +\varepsilon^2}
 \chi(\nn,\nn') \,dX\,dX' \nonumber \\
 & = \int \frac{q_\lambda(R_{\nn'}x,x')}{\tilde{\sigma}_{[\lambda]}^{(\mathcal{L})} + {\sf s}(\nn,\overset{\circ}{\nn}{}') + 2i\varepsilon t_{[\lambda]} +\varepsilon^2}
 \chi(R_{\nn'}\nn,\nn')\,dX \,dX' \,.
\end{align}
Using the standard spherical angular coordinates $(\vartheta,\varphi) = \nn$, with $\vartheta = 0$ corresponding to the ``north pole point''
$ = \overset{\circ}{\nn}{}'$, the half of the squared geodesic distance on the sphere with radius $r_*$ takes the simple form 
\begin{align} \label{eq:entry}
 {\sf s}(\vartheta,\varphi,\overset{\circ}{\nn}{}') = \frac{r_*^2\vartheta^2}{2}\,,
\end{align}
and we thus obtain, on writing $\xi_\lambda(x,x') = \chi(R_{\nn'}\nn,\nn') q_\lambda(R_{\nn'}x,x')$,
\begin{align}
 & \int \frac{q_\lambda(R_{\nn'}x,x')}{\tilde{\sigma}_{[\lambda]}^{(\mathcal{L})} + {\sf s}(\nn,\overset{\circ}{\nn}{}') + 2i\varepsilon t_{[\lambda]} +\varepsilon^2}
 \chi(R_{\nn'}\nn,\nn')\,dX \,dX'  \nonumber \\
 & = \int \int_{\vartheta = 0}^\delta \frac{\xi_\lambda(x,x')}{\tilde{\sigma}_{[\lambda]}^{(\mathcal{L})} + r_*^2\vartheta^2/2 + 2i\varepsilon t_{[\lambda]} +\varepsilon^2}
 \sin(\vartheta)d\vartheta \,d\varphi\,dU\,dV\, \,dX'
\end{align}
since in the polar coordinates chosen, $r_*^2\vartheta^2/2$ is the half of the squared geodesic distance between $\overset{\circ}{\nn}{}'$ and $\nn$ on 
the sphere with radius $r_*$, and $\xi_\lambda(x,x') = 0$ if $\vartheta > \delta$ by the properties of $\chi$. Now carrying out a partial 
integration with respect to $\vartheta$ and observing
\begin{align}
 \frac{1}{\tilde{\sigma}^{(\mathcal{L})}_{[\lambda]} + r_*^2\vartheta^2/2 + 2i \varepsilon t_{[\lambda]} + \varepsilon^2}
 = \frac{1}{r_*^2\vartheta} \partial_\vartheta 
 \ln(\tilde{\sigma}^{(\mathcal{L})}_{[\lambda]} + r_*^2\vartheta^2/2 + 2i \varepsilon t_{[\lambda]} + \varepsilon^2) \,,
\end{align}
we are led to
\begin{align}
 &\int \int_{\vartheta = 0}^\delta \frac{\xi_\lambda(x,x')}{\tilde{\sigma}^{(\mathcal{L})}_{[\lambda]} + r_*^2\vartheta^2/2 - 2i \varepsilon t_{[\lambda]} + \varepsilon^2} 
  \sin(\vartheta)d\vartheta \,d\varphi\,dU\,dV\,dX' \nonumber  \\ \label{eq:rest}
  & = \int \left[ \ln(\tilde{\sigma}^{(\mathcal{L})}_{[\lambda]} + r_*^2\vartheta^2/2 + 2i \varepsilon t_{[\lambda]} + \varepsilon^2)
 \xi_\lambda(x,x') \frac{{\rm sinc}(\vartheta)}{r_*^2} \right]_{\vartheta = 0}^\delta \,d\varphi\,dU\,dV\,dX'   \\
& - \int \int_{\vartheta = 0}^\delta \ln(\tilde{\sigma}^{(\mathcal{L})}_{[\lambda]} + r_*^2\vartheta^2/2 + 2i \varepsilon t_{[\lambda]} + \varepsilon^2)\,
\partial_\vartheta \left( \xi_\lambda(x,x') \frac{{\rm sinc}(\vartheta)}{r_*^2}\right) \, d\vartheta \, d\varphi\, dU \,dV \,dX'\,. \label{eq:dtheta}
\end{align}
We note that 
\begin{align}
 \ln(\tilde{\sigma}_\varepsilon(x,x')) = & \ln| \tilde{\sigma}^{(\mathcal{L})}(U,V,U',V') + {\sf s}(\nn,\nn') + 2i\varepsilon t(x,x')  + \varepsilon^2|  \nonumber \\
 & + i {\rm arg}(\tilde{\sigma}^{(\mathcal{L})}(U,V,U',V') + {\sf s}(\nn,\nn') + 2i\varepsilon t(x,x') +  \varepsilon^2)
\end{align}
so that Lemma \ref{lm:delta-estimate} applies to the expression in \eqref{eq:dtheta}, with $\alpha = (\lambda,\varepsilon) \in \mathbb{R}^2$\,, on noting that
the 
argument function part stays uniformly bounded in $(\lambda,\varepsilon)$, resulting in a contribution in \eqref{eq:dtheta} which is $O(\delta)$ as $\delta \to 0$.
Therefore,
supposing that $\lambda$ and $\varepsilon$ have been chosen sufficiently small, and likewise that $\delta > 0$ is sufficiently small, we can conclude that  
{\small
\begin{align}
 \sup_{\lambda,\varepsilon}\,\left|\int \int_{\vartheta = 0}^\delta \ln(\tilde{\sigma}^{(\mathcal{L})}_{[\lambda]} + r_*^2\vartheta^2/2 - 2i \varepsilon t_{[\lambda]} + \varepsilon^2)\,
\partial_\vartheta \left( \xi_\lambda(x,x') \frac{{\rm sinc}(\vartheta)}{r_*^2}\right) \, d\vartheta \, d\varphi\, dU \,dV \,dX' \,\right| \le \kappa_1 \delta^{1/3} 
\end{align}}
with a suitable positive constant $\kappa_1$. 

In a similar manner we find, provided that $\lambda,\varepsilon$ and $\delta$ are sufficiently close to 0, on account of Lemma \ref{lm:delta-estimate}
\begin{align}
 & \sup_{\lambda,\varepsilon} \left| \int \ln(\tilde{\sigma}_\varepsilon(x_\lambda,x'_\lambda)) \tilde{Y}(x_\lambda,x'_\lambda) \chi(\nn,\nn') F(x)F(x') P(\lambda U,V,\lambda U',V')\,dX\,dX' \right| \\
 \le &  \sup_{\lambda,\varepsilon}\, \left| \int \int_{\vartheta = 0}^\delta \ln(\tilde{\sigma}^{(\mathcal{L})}_{[\lambda]} + r_*^2\vartheta^2/2 + 2i \varepsilon t_{[\lambda]} + \varepsilon^2)
 k_\lambda(x,x') \,d\vartheta \, d\varphi\, dU \,dV \,dX'\, \right| 
 \le \kappa_2 \delta^{1/3} \nonumber
\end{align}
with a family of smooth functions $k_\lambda(x,x')$ of $x$ and $x'$  which is uniformly bounded and uniformly compactly supported in $\lambda$; $\kappa_2 > 0$ is a suitable constant. 

Furthermore, since $\tilde{Z}(x,x')$ is $C^\infty$, we see that 
\begin{align}
 \sup_{\lambda,\varepsilon}\,\left| \int \tilde{Z}(x,x')\chi(\nn,\nn') F_\lambda(x)F_\lambda(x')  P(U,V,U',V')r_*^2\,dX\,dX' \right| = O(\delta)
\end{align}
if $\lambda$ and $\delta$ are small enough. 

Summarizing our findings up to this point, we see that, on choosing
\begin{align}
 q_\lambda(x,x') = \tilde{\psi}(x_\lambda,x'_\lambda)\frac{\tilde{\Delta}^{1/2}(x_\lambda,x'_\lambda)}{8 \pi^2}F(x)F(x')P(\lambda U,V,\lambda U',V')
\end{align}
in \eqref{eq:qlambda}, we obtain 
\begin{align}
 \lim_{\lambda \to 0+}\, &  w^{(2)}(F_\lambda,F'_\lambda) \\
  & =  \lim_{\lambda \to 0+}\,  \lim_{\varepsilon \to 0+} \int \left[ \ln(\tilde{\sigma}^{(\mathcal{L})}_{[\lambda]} + r_*^2\vartheta^2/2 + 2i \varepsilon t_{[\lambda]} + \varepsilon^2)
 \xi_\lambda(x,x') \frac{{\rm sinc}(\vartheta)}{r_*^2} \right]_{\vartheta = 0}^\delta \,d\varphi\,dU\,dV\,dX' 
 \nonumber \\ 
 & + O(\delta^{1/3}) \nonumber
\end{align}
for any sufficiently small $\delta > 0$.
However, the integral expression is independent of $\delta$: Recalling that $\xi_\lambda(x,x') = 0$ if $\vartheta > \delta$, the evaluation of the integral expression at  $\vartheta = \delta$ 
vanishes, and the resulting expression, as we will see, is independent of $\chi_\delta$ which is contained in the definition of $\xi_\lambda(x,x')$. Therefore, since $\delta$ may be chosen 
arbitrarily small, we now obtain 
\begin{align}
 & \lim_{\lambda \to 0+}\,  w^{(2)}(F_\lambda,F'_\lambda) \\
  & =  \lim_{\lambda \to 0+}\,  \lim_{\varepsilon \to 0+} \int - \left. \ln(\tilde{\sigma}^{(\mathcal{L})}_{[\lambda]} + r_*^2\vartheta^2/2 + 2i \varepsilon t_{[\lambda]} + \varepsilon^2)
 \xi_\lambda(x,x') \frac{{\rm sinc}(\vartheta)}{r_*^2} \right|_{\vartheta = 0} \,d\varphi\,dU\,dV\,dX' \,. \nonumber
\end{align}
Evaluating the integral expression at $\vartheta = 0$, observing $\xi_\lambda(x,x') = \chi(R_{\nn'}\nn,\nn') q_\lambda(R_{\nn'}x,x')$, results in 
\begin{align}
 & \lim_{\lambda \to 0+}\,   w^{(2)}(F_\lambda,F'_\lambda)  \label{eq:remaining-expr}\\
& =    \lim_{\lambda \to 0+}\,  \lim_{\varepsilon \to 0+} \int  -\ln(\tilde{\sigma}^{(\mathcal{L})}_{[\lambda]} + 2i \varepsilon t_{[\lambda]} + \varepsilon^2) q_\lambda(U,V,\nn',U'V',\nn')  \frac{2\pi}{r_*^2}
   \, dU\,dV\, dU'\, dV' \,d\Omega^2(\nn')   \nonumber 
\end{align}
To see this, note first that, in the coordinates chosen, $x|_{\vartheta = 0} = (U,V,\overset{\circ}{\nn}{}')$, implying ${\sf s}(R_{\nn'}\nn,\nn') =  {\sf s}(\nn,\overset{\circ}{\nn}{}') = {\sf s}(\overset{\circ}{\nn}{}',
\overset{\circ}{\nn}{}')= 0$ 
and therefore, $\chi(R_{\nn'}\nn,\nn') = 1$.
On the other hand, $\nn =  \overset{\circ}{\nn}{}'$ also means that there is no $\varphi$-dependence in the integrand, and the integral with respect to $\varphi$ can be carried out,
just contributing a factor $2\pi$. Moreover, it implies that $q_\lambda(R_{\nn'}x,x') = q_\lambda(U,V,\nn',U',V',\nn')$ in the last integral. 
\\[6pt]
We are thus left with having to evaluate the limits of the right hand side in \eqref{eq:remaining-expr}. We note that 
\begin{align}
 \left.\tilde{\sigma}_{[\lambda]}^{(\mathcal{L})}\right|_{\lambda = 0} = \left. \tilde{\sigma}^{(\mathcal{L})}(\lambda U,V,\lambda U',V')
 \right|_{\lambda =0} = 0 
\end{align}
and therefore, the Taylor expansion of $\tilde{\sigma}_{[\lambda]}^{(\mathcal{L})}$ in $\lambda$ at $\lambda = 0$ up to second order 
yields 
\begin{align}
 \tilde{\sigma}_{[\lambda]}^{(\mathcal{L})} = \lambda (U - U')(\tilde{V} - \tilde{V}') + R_\lambda(U,V,U',V') 
\end{align}
with $R_\lambda = O(\lambda^2)$ uniformly on compact sets in $U,U'$ and $V,V'$ while $\tilde{V}$ (and $\tilde{V}'$) is a geodesic 
parameter of the lightlike curves $V \mapsto (0,V,\nn)$ with respect to the conformally transformed metric $\tilde{g}_{ab}$ chosen
such that $dU_a = \tilde{g}_{ab} (\partial/\partial \tilde{V})^b$. This can be seen from eqns.\ (3.3) and (3.4) in \cite{Poisson}; note 
that $V$ is an affine parameter for the said lightlike curves with respect to $g_{ab}$ but not necessarily with respect to the conformally 
transformed metric $\tilde{g}_{ab}$. Using the form of the ``Lorentzian'' part of the conformally transformed metric
\begin{align}
 -2 \eta^2(U,V)A(U,V)dUdV \,,
\end{align}
it is not difficult to check that $\tilde{V} = \tilde{V}(V)$ has the property 
\begin{align}
 \tilde{V} - \tilde{V}' = \int_V^{V'} \eta^2(0,V_1)\,dV_1 \quad \text{implying} \quad 
 \tilde{V} - \tilde{V}' = (V - V')\gamma(V,V')
\end{align}
with a positive, jointly continuous function $\gamma(V,V')$ where $\gamma(V,V')$ and $1/\gamma(V,V')$ are bounded 
when $V$ and $V'$ range over compact sets. Consequently, we have that 
{\small
\begin{align}
 & \lim_{\lambda \to 0+}\,   w^{(2)}(F_\lambda,F'_\lambda)  \nonumber\\
& =    \lim_{\lambda \to 0+}\,  \lim_{\varepsilon \to 0+} \int - \ln(
\lambda(U - U')(V - V') \gamma(V,V') + R_\lambda(U,V,U',V') + 2i \varepsilon t_{[\lambda]} + \varepsilon^2) 
\times  \nonumber \\
& \quad \quad \quad \quad \quad \times \ 
q_\lambda(U,V,\nn',U'V',\nn')  \frac{2\pi}{r_*^2}
   \, dU\,dV\, dU'\, dV' \,d\Omega^2(\nn')  \nonumber \\
   & = \lim_{\lambda \to 0+}\,  \lim_{\varepsilon \to 0+} \int - [ \ln(\lambda \gamma)
    + \ln((U - U')(V - V') + \varrho_\lambda(U,V,U',V') +\lambda^{-1} \gamma^{-1}[2i \varepsilon t_{[\lambda]} + \varepsilon^2])]
    \times  \nonumber   \\
& \quad \quad \quad \quad \quad \times \ 
q_\lambda(U,V,\nn',U'V',\nn')  \frac{2\pi}{r_*^2}
   \, dU\,dV\, dU'\, dV' \,d\Omega^2(\nn') \label{eq:ln-parts}
\end{align} }
with $\varrho_\lambda = \lambda^{-1}\gamma^{-1}R_\lambda$,
and we have abbreviated $\gamma = \gamma(V,V')$.

Let us first consider the 
$\varepsilon$-independent part in \eqref{eq:ln-parts}. We split $\ln(\lambda \gamma) = \ln(\lambda) + \ln(\gamma)$. Then the 
limits can be carried out to yield\footnote{We reinsert the abbreviation $dX' = dU'\, dV' \,d\Omega^2(\nn')$}
\begin{align}
 \lim_{\lambda \to 0+}\, \int -\ln(\gamma(V,V')) q_\lambda(U,V,\nn',U',V',\nn') \frac{2\pi}{r_*^2}
   \, dU\,dV\, dX' = 0
\end{align}
because the only $U$ and $U'$ dependence in $q_0(U,V,U',V')$ comes from $\partial_Uf(U)$ and $\partial_{U'}f'(U')$ and therefore,
since $f$ and $f'$ are compactly supported, we find that the resulting integral vanishes.
On the other hand, we also have $q_\lambda(U,V,\nn',U',V',\nn') = q_0(U,V,\nn',U',V',\nn') + q^{(1)}_\lambda(U,V,U',V';\nn')$ where $q^{(1)}_\lambda(U,V,U',V';\nn') = O(\lambda)$ uniformly on compact sets in $U,U'$ and $V,V'$. Then, 
\begin{align}
\lim_{\lambda \to 0+}\, \int \ln(\lambda) (q_0(U,V,\nn',U',V',\nn') + q^{(1)}_\lambda(U,V,U',V';\nn')) \frac{2\pi}{r_*^2}
   \, dU\,dV\, dX' = 0
\end{align}
since, as in the previous argument, $q_0(U,V,\nn',U',V',\nn')$ depends on $U$ and $U'$ only through $\partial_U f$ and $\partial_{U'} f'$, and 
$\ln(\lambda)q_\lambda^{(1)}(U,V,U',V';\nn') \to 0$ as $\lambda \to 0$ uniformly on compact sets.

Therefore, setting $\beta_{\lambda,\varepsilon}(U,V,U',V') = \varrho_\lambda(U,V,U',V')+  \gamma^{-1}[2i \varepsilon t_{[\lambda]} + \lambda \varepsilon^2]$, we obtain 
\begin{align}
 \lim_{\lambda \to 0+}\,   w^{(2)}(F_\lambda,F'_\lambda) 
& =    \lim_{\lambda \to 0+}\,  \lim_{\varepsilon \to 0+} \int - \ln(
(U - U')(V - V') + \beta_{\lambda,\varepsilon}(U,V,U',V')) \times \nonumber \\
& \quad \quad \quad \quad \quad  \times q_\lambda(U,V,\nn',U',V',\nn')\frac{2\pi}{r_*^2}
   \, dU\,dV\, dX' 
\end{align}
where we have made use of the fact that, owing to the limit $\varepsilon \to 0+$ being taken prior to $\lambda \to 0+$, one 
may redefine $\varepsilon$ as $\varepsilon/\lambda$ without changing the result of the limits. 

Then we choose $\delta > 0$ and split the integration over the $U,U'$ and $V,V'$ coordinates into the domains
\begin{align}
  D_{>}(\delta) = |(U - U')(V -V')| \ge \delta^2 \,, \quad \ \ D_{<}(\delta) =|(U - U')(V - V')| < \delta^2 \,.
\end{align}
Furthermore, we change from the $(U,V)$ coordinates to $(T,Y)$ coordinates where 
\begin{align}
 T = \frac{1}{2}(U + V) \,, \quad \quad Y = \frac{1}{2}(U - V) 
\end{align}
and using a coordinate substitution $\bar{T} = T - T'$, $\bar{Y} = Y - Y'$, one arrives at 
{\small
\begin{align}
& \int_{D_{<}(\delta) } \ln(
(U - U')(V - V') + \beta_{\lambda,\varepsilon}(U,V,U',V'))  q_\lambda(U,V,\nn',U',V',\nn')\frac{2\pi}{r_*^2}
   \, dU\,dV\, dX' \\
   & = \int \int_{4| \bar{T}^2 - \bar{Y}^2| < \delta^2} \ln( \bar{T}^2 - \bar{Y}^2 + \bar{\beta}_{\lambda,\varepsilon}(\bar{T},\bar{Y},
   T',Y')) \bar{q}_\lambda(\bar{T},\bar{Y},T',Y';\nn')\frac{2\pi}{r_*^2} \, d\bar{T}\,d\bar{Y}\,dT'\,dY'\,d\Omega^2(\nn') \nonumber
\end{align}
}
where $\bar{\beta}_{\lambda,\varepsilon}$ and $\bar{q}_\lambda$ are the $\bar{T},\bar{Y},T',Y'$-coordinate versions of 
$\beta_{\lambda,\varepsilon}$ and $q_\lambda$. We see that the right hand side is in a form to which Part {\bf (B)} of 
Lemma \ref{lm:delta-estimate} applies (with $\alpha = (\lambda,\varepsilon)$), and arguing analogously as with \eqref{eq:dtheta} before, we conclude that,
once $\varepsilon,\lambda$ and $\delta$ have been chosen sufficiently small, it holds that 
{\small
\begin{align}
 \sup_{\lambda,\varepsilon}&\left|\int_{D_{<}(\delta)}  \ln(
(U - U')(V - V') + \beta_{\lambda,\varepsilon}(U,V,U',V'))  q_\lambda(U,V,\nn',U',V',\nn')\frac{2\pi}{r_*^2}
   \, dU\,dV\, dX' \right| \nonumber \\ 
   & = O(\delta^{1/3}) 
\end{align} }
allowing to conclude
\begin{align}
 \lim_{\lambda \to 0+}\,   w^{(2)}(F_\lambda,F'_\lambda) 
& =  \lim_{\delta \to 0}\,  \lim_{\lambda \to 0+}\,  \lim_{\varepsilon \to 0+} \int_{D_{>}(\delta)}  \ln(
(U - U')(V - V') + \beta_{\lambda,\varepsilon}(U,V,U',V')) \times \nonumber \\
& \quad \quad \quad \quad  \times q_\lambda(U,V,\nn',U',V',\nn')\frac{2\pi}{r_*^2}
   \, dU\,dV\, dX' 
\end{align}
The limits can now be carried out performing the limit $\varepsilon \to 0$ followed by $\lambda \to 0$ prior to integration,
to yield 
\begin{align}
\lim_{\lambda \to 0+}\,   w^{(2)}(F_\lambda,F'_\lambda) =
 \int [\ln(|(U - U')|) - i\pi \theta(U - U')] q_0(U,V,\nn',U',V',\nn') \frac{2\pi}{r_*^2}
   \, dU\,dV\, dX' \label{eq:almost-there}
\end{align}
where $\theta$ denotes the Heaviside function. 
To see this, we note that 
{\small
\begin{align}
 & \lim_{\lambda \to 0+} \lim_{\varepsilon \to 0+}
  \ln(
(U - U')(V - V') + \beta_{\lambda,\varepsilon}(U,V,U',V'))  q_\lambda(U,V,\nn',U',V',\nn') \\
& = [\ln(|(U - U')(V - V')|) + i \pi \theta((U' - U)(V - V')){\rm sign}(\gamma^{-1}(V- V'))] q_0(U,V,\nn',U',V',\nn') \nonumber
\end{align} }
Using $\ln(|(U - U')(V - V')|) = \ln|U - U'| + \ln|V - V'|$, one can see again that the $\ln|V -V'|$ term gives a vanishing 
contribution on integration with respect to $U$ and $U'$ because $q_0(U,V,\nn',U',V',\nn')$ depends on $U$ and $U'$ only through
$\partial_Uf$ and $\partial_{U'} f'$. Moreover, since $\gamma^{-1} = \gamma^{-1}(V,V') > 0$, we have 
$\theta((U' - U)(V - V')){\rm sign}( \gamma^{-1}(V- V')) = \theta((U' - U){\rm sign}(V - V')){\rm sign}(V - V')$;
and since 
\begin{align}
 \int[ \theta(U - U') + \theta(U' - U)] \partial_U f(U,V,\nn') \partial_{U'}f'(U',V',\nn') dU \,dU'  = 0
\end{align}
we can conclude that 
\begin{align}
 & \int i \pi \theta((U' - U)(V - V')){\rm sign}(\gamma^{-1}(V- V')) q_0(U,V,\nn',U',V',\nn') \,dU\,dV\,dX' \nonumber \\
 & = - \int i \pi \theta(U- U') q_0(U,V,\nn',U',V',\nn') \,dU\,dV\,dX'
 \end{align}
showing that \eqref{eq:almost-there} holds.

It is well-known -- or can easily be derived by arguments analogous to those given in the proof up to now, using partial integrations with respect to
$U$ and $U'$ -- that 
\begin{align} 
&\int [\ln(|(U - U')|) - i\pi \theta(U - U')] q_0(U,V,\nn',U',V',\nn') \frac{2\pi}{r_*^2}
   \, dU\,dV\, dX' \nonumber \\
  & = \lim_{\varepsilon \to 0+} \int \frac{ q_0(U,V,\nn',U',V',\nn') }{(U - U' + i\varepsilon)^2} \frac{2\pi}{r_*^2}
   \, dU\,dV\, dX'\,.
\end{align}
Finally, we observe that $\tilde{\psi}(0,V,\nn',0,V',\nn') = 1$ because the points $(0,V,\nn')$ and $(0,V',\nn')$ are causally related 
and lie, by assumption, in a causal normal neighbourhood. Therefore, we have shown that 
\begin{align}
 \lim_{\lambda \to 0+}\,   w^{(2)}(F_\lambda,F'_\lambda)  & =
  \lim_{\varepsilon \to 0+} \int \frac{ q_0(U,V,\nn',U',V',\nn') }{(U - U' + i\varepsilon)^2} \frac{2\pi}{r_*^2}
   \, dU\,dV\, dX' \\
 & =  -\frac{1}{r_*^2{\pi}} \int  \frac{ f(U,V,\nn) f'(U',V',\nn)}{(U-U'-i\varepsilon)^2}Q(V,V',\nn)\,
dU\,dU\,'dV\,dV'\,d\Omega^2(\nn) \nonumber
\end{align}
as claimed in the statement (I) of the Theorem. 
\\[6pt]
The second statement is easily proved since the $\mu \to 0$ limit can be taken directly as the distribution $L$ only involves 
integrations against continuous, bounded functions with respect to the $V$ and $V'$ coordinates.
In fact, the limits $\varepsilon \to 0$ in the definition of $L$ (resp., $\Lambda$) can be interchanged with the 
limit $\mu \to 0$ used to pass from $L$ to $\Lambda$. 
The result claimed in (II) then follows on observing that the square root of the van Vleck-Morette determinant equals 
the unit at coinciding points, so $\tilde{\Delta}^{1/2}(0,0,\nn,0,0,\nn) = 1$, and $\eta^{-1}(0,0) = 1$ as well as 
$A(0,0) = 1$.

This completes the proof of the Theorem. \hfill $\Box$


\end{document}